\documentclass[twocolumn]{aastex6}
\usepackage[yyyymmdd]{datetime}
\usepackage{bm}
\usepackage{mathbbol}
\usepackage{url}
\usepackage{amsmath}
\usepackage{algpseudocode}
\usepackage[ruled]{algorithm}

\DeclareMathOperator{\argmin}{arg\,min}

\begin{document}

\title{LINEAR: A Novel Algorithm for Reconstructing Slitless Spectroscopy from HST/WFC3\footnote{Part of this work carried out at Space Telescope Science Institute (STScI) was performed in the context of Science Operations Center contracts for the HST, and WFIRST missions funded by NASA Goddard Space Flight Center.  Based on observations made with the NASA/ESA Hubble Space Telescope, obtained from the data archive at the Space Telescope Science Institute. STScI is operated by the Association of Universities for Research in Astronomy, Inc. under NASA contract NAS 5-26555. }}

\author{R.~E.~Ryan Jr., S.~Casertano,  N.~Pirzkal}
\affil{Space Telescope Science Institute\\
3700 San Martin Ave.\\
Baltimore, MD 21210, USA}

\newcommand{\LINEAR}{\texttt{LINEAR}}
\newcommand{\URL}{http://}
\newcommand{\aXe}{\texttt{aXe}}
\newcommand{\ith}{$i^\mathrm{th}$}
\newcommand{\jth}{$j^\mathrm{th}$}
\newcommand{\ie}{\textit{i.e.}}
\newcommand{\eg}{\textit{e.g.}}

\newcommand{\Input}{\textbf{input:}}
\newcommand{\each}{\textbf{each:}}
\newcommand{\Output}{\textbf{output:}}

\newcommand{\Sect}[1]{Section~\ref{#1}}
\newcommand{\Eqn}[1]{equation~(\ref{#1})}
\newcommand{\Eqns}[2]{equations~(\ref{#1}) and (\ref{#2})}
\newcommand{\EQN}[1]{Equation~(\ref{#1})}
\newcommand{\Fig}[1]{Figure~\ref{#1}}
\newcommand{\Tab}[1]{Table~\ref{#1}}

\begin{abstract}

We present a grism extraction package (\LINEAR) designed to
reconstruct one-dimensional spectra from a collection of slitless
spectroscopic images, ideally taken at a variety of orientations,
dispersion directions, and/or dither positions.  Our approach is to
enumerate every transformation between all direct image positions
(ie.~a potential source) and the collection of grism images at all
relevant wavelengths.  This leads to solving a large, sparse system of
linear equations, which we invert using the standard LSQR algorithm.
We implement a number of color and geometric corrections (such as flat
field, pixel-area map, source morphology, and spectral bandwidth), but
assume many effects have been calibrated out (such as basic
reductions, background subtraction, and astrometric refinement).  We
demonstrate the power of our approach with several Monte Carlo
simulations and the analysis of archival data.  The simulations
include astrometric and photometric uncertainties, sky-background
estimation, and signal-to-noise calculations.  The data are G141
observations obtained with the Wide-Field Camera 3 of the Hubble
Ultra-Deep Field, and show the power of our formalism by improving the
spectral resolution without sacrificing the signal-to-noise (a
tradeoff that is often made by current approaches).  Additionally, our
approach naturally accounts for source contamination, which is only
handled heuristically by present softwares.  We conclude with a
discussion of various observations where our approach will provide
much improved spectral one-dimensional spectra, such as crowded fields
(star or galaxy clusters), spatially resolved spectroscopy, or surveys
with strict completeness requirements.  At present our software is
heavily geared for Wide-Field Camera~3 IR, however we plan extend the
codebase for additional instruments.

\end{abstract}


\section{Introduction} \label{sec:intro}

Slitless spectroscopy uses a dispersive  and transmissive element in a
collimated  beam to  produce  a  dispersed image  of  the sky.   Since
slitless  spectroscopy  lacks  a  physical aperture  to  restrict  the
spatial  extent of  the incoming  light, the  observed two-dimensional
spectra can  be thought  of as  a convolution of  the spectrum  of the
source,  as  processed by  the  dispersing  element, and  its  spatial
profile.   Therefore the  physical  extent of  the  sources along  the
dispersion  axis will  limit  the achievable  spectral resolution,  in
analogy  to  the relationship  between  slit-width  and resolution  in
traditional,  aperture-based spectroscopy.   However,  the  lack of  a
physical  aperture   is  also   the  primary  strength   for  slitless
spectroscopy, since  one can obtain  a complete spectroscopic  view of
all the sources  in the field, free from  major observational planning
or  targeting considerations  (such as  slit masks,  fiber placements,
\emph{etc.}).

The  complete  multiplexing  of   slitless  spectroscopy  leads  to  a
significant  data  processing  challenge: how  to  handle  overlapping
dispersions  and  source  contamination \citep[\eg][]{axe}.   With  an
optical  model  of  the  detector, it  is  often  straight-forward  to
identify the pixels in a  dispersed image that have contributions from
multiple  objects  or  spectral  orders, however  how  to  treat  such
contaminated pixels is far less  obvious.  One approach for extracting
one-dimensional  spectra  estimates  the contamination  based  on  the
available     broadband    colors     \citep[\eg][]{pirz09,brammer12}.
Alternatively  one  may opt  to  model  the  dispersed images  with  a
parametric     \citep[\eg][]{bc03}      or     non-parametric     data
\citep[\eg][]{pickles},   which  will   naturally   account  for   the
overlapping  spectra.   While  this approach  can  directly  transform
observations  into  concrete,  astrophysical  answers,  it  is  deeply
dependent on  the assumptions  of the spectral  models.  On  the other
hand, \citet{barger12} present a  non-parameteric method for combining
data  from  multiple  orients  from  slitless  spectroscopy  from  the
\emph{Galaxy   Evolution  Explorer}   \citep[\emph{GALEX};][]{martin}.
Their approach relies on certain assumptions (\eg\ unresolved sources,
minimal  contamination, and  $\sim\!10-100$  orients), which  normally
only apply to typical observations.

With  these considerations,  we  propose an  alternative approach  for
extracting  one-dimensional  spectra  from a  set  of  two-dimensional
images with  a singular question  in mind: \emph{What is  the optimal,
  non-parametric  spectrum  for  each object  that  is  simultaneously
  consistent  with the  available slitless  spectroscopy?} We  develop
this as  an ``optimal''  strategy, as  it minimizes  a goodness-of-fit
parameter for an  overconstrained problem. This work  presents our new
approach for  extracting slitless spectroscopy,  particularly tailored
to addressing source contamination (or confusion).

This paper  is structured as  follows:~in \Sect{sec:primer} we  give a
brief  primer on  slitless  spectroscopy,  in \Sect{sec:algorithm}  we
describe  our new  algorithm, in  \Sect{sec:tests} we  present several
tests of  our approach  and place  constraints on  the quality  of the
calibrations, in  \Sect{sec:hudf} we use  the G141 data in  the Hubble
Ultra-Deep  Field  as a  case  study,  in \Sect{sec:mods}  we  outline
several  key   improvements  we   consider  for  later   releases,  in
\Sect{sec:discuss} we discuss several  applications where we foresee a
unique role  for \LINEAR,  and in \Sect{sec:summary}  we give  a brief
summary.   We  include a  glossary  in  the  Appendix to  clarify  any
confusion with  notation.  We  quote all magnitudes  in the  AB system
\citep{og}.

\section{Grism Primer}\label{sec:primer}

We wish the review several  key aspects of the observation, reduction,
and analysis  of slitless spectroscopy,  as the concepts  and notation
are  important to  frame  our  methodology. For  clarity,  all of  the
variables given in calligraphic font are two-dimensional images.
To support slitless spectroscopic  observations, many additional forms
of data are  required and/or concurrently collected.   We define these
components as:
\begin{description}
  \item[dispersed imaging] the two-dimensional spectroscopic data.  We
    may also refer to these as ``grism'' or slitless
    data/observations/imaging.  

  \item[pre-imaging]  standard  imaging  taken concurrently  with  the
    dispersed  imaging.  These  data  are often,  but not  exclusively
    \citep[\eg][]{pirzkal_isr},  taken in  a broadband  whose response
    closely matches the grism response to astrometrically register the
    dispersed images.  As such, these data generally have much shorter
    exposure   times   than   the  primary   grism   data.    Recently
    \citet{bohlin15}  describe  options  for  using  the  zeroth-order
    dispersion for the reference position, which obviates the need for
    the pre-imaging, but we do not consider any such options.

  \item[direct  imaging] existing  imaging,  analogous  to a  ``finding
    chart.''   The  nature  of   these  data  differ  greatly  between
    programs, but are often  used for source identification (discussed
    more below).   These images  also often  serve as  the astrometric
    reference for the aforementioned pre-imaging.

\end{description}

The undispersed  position of each  source (\ie, the  detector position
the source  element would have  had in  the absence of  the dispersive
element) is given as $(x_0,y_0)$.  For much of this work, we decompose
the source into  direct image pixels, and trace the  flux in each such
pixel by transforming the positions of  the four corners as a function
of  wavelength.   For some  instruments,  such  as HST/WFC3  with  the
infrared grisms, the spectral trace at any location in the detector is
close  to  parallel  to  one  pixel axis  (often  the  $x$-axis),  the
curvature  of the  trace is  very slight,  and the  wavelength can  be
represented as  a low-order polynomial  of the displacement  along the
trace.  Under these  conditions, a simple representation  of the trace
and  dispersion solution  can be  used, as  follows. This  undispersed
position  is  used to  define  a  polynomial  for the  spectral  trace
($\tilde{x},\tilde{y}$):
\begin{equation}\label{eqn:trace}
\tilde{y}(\tilde{x})=\sum_{i=0}^{n}\alpha_i(x_0,y_0)\,\tilde{x}^i,
\end{equation}
where the coefficients $\{\alpha\}$ depend on the source position (see
\Eqn{eqn:alpha}  below).   The  spectral  trace is  transformed  to  a
detector position $(x,y)$ by adding appropriate offsets:
\begin{eqnarray}\label{eqn:offset}
x &=& x_0 + x_\mathrm{off}(x_0,y_0) + \tilde{x},\\
y &=& y_0 + y_\mathrm{off}(x_0,y_0) + \tilde{y}(\tilde{x}).
\end{eqnarray}
The coordinate  offsets ($x_\mathrm{off},y_\mathrm{off}$)  also depend
on the source positions (given by \Eqns{eqn:x0}{eqn:y0}).

The wavelength is given as a  function of path length along the trace,
generally assumed to be a polynomial
\begin{equation}\label{eqn:disp}
\lambda(s)=\sum_{i=0}^{n}\beta_i(x_0,y_0)\,s^i,
\end{equation}
and of course the path length has the usual form:
\begin{equation}\label{eqn:path}
s(\tilde{x})=\int\limits_0^{\tilde{x}}\!\sqrt{1+\left(\frac{\mathrm{d}\tilde{y}}{\mathrm{d}\tilde{x}'}\right)^2}\,\mathrm{d}\tilde{x}'.
\end{equation}
The coefficients  $\{\beta\}$ are again given  by \Eqn{eqn:beta}.  In
the special case of WFC3/IR, the  spectral trace is linear in detector
position,  and   so  this  integral  can   be  computed  analytically.
Currently, \LINEAR\ is capable of  dealing with first- or second-order
spectral        traces,         which        excludes        WFC3/UVIS
\citep[see][]{kunt09,pirz17}.  It  is  straight-forward to  extend  to
higher-order  spectral traces,  however  this will  require solving  a
Volterra  equation of  the first  kind.  The  parameterization of  the
wavelength  as a  function of  the path-length  represents a  specific
choice  made   in  the  calibration   of  the  WFC3  grisms   and  our
implementation here.

As  mentioned  above,  the  field-dependence of  the  spectral  trace,
dispersion,  and spatial offsets  ($x_\mathrm{off},y_\mathrm{off}$) is
described as  two-dimensional polynomials that are similar  in form to
the simple-imaging polynomials \citep[SIP;][]{shupe}
\begin{eqnarray} 
  \label{eqn:alpha}
  \alpha_i(x_0,y_0)&=&\sum_{j=0}^{n}\sum_{k=0}^{n}\alpha_{i,j,k}\,{x_0}^j\,{y_0}^k;\,j+k\leq n\\
  \label{eqn:beta}
  \beta_i(x_0,y_0)&=&\sum_{j=0}^{n'}\sum_{k=0}^{n'}\beta_{i,j,k}\,{x_0}^j\,{y_0}^k;\,j+k\leq n'\\
  \label{eqn:x0}
  x_\mathrm{off}(x_0,y_0)&=&\sum_{j=0}^m\sum_{k=0}^{m}\xi_{j,k}\,{x_0}^j\,{y_0}^k;\,j+k\leq m\\
  \label{eqn:y0}
  y_\mathrm{off}(x_0,y_0)&=&\sum_{j=0}^{m'}\sum_{k=0}^{m'}\eta_{j,k}\,{x_0}^j\,{y_0}^k;\,j+k\leq m' 
\end{eqnarray}
The    \aXe-based    reference    files   store    the    coefficients
$\alpha_{i,j,k}$, $\beta_{i,j,k}$, $\xi_{j,k}$, and $\eta_{j,k}$ with
a     single    index     that    combines     $(j,k)\!\rightarrow\!j$
\citep[\eg][]{pirzkal_isr}. 

A more  general representation of  the trace and  dispersion solution,
suitable for  traces that  have significant  curvature or  become more
closely  aligned  with  the  $y$-axis   of  the  detector  and/or  the
wavelength  equation  is  more  complex,  may  rely  on  a  parametric
representation of the trace and wavelength as:
\begin{eqnarray}\label{eqn:genrep}
  \tilde{x}(t) &=& \alpha'(x_0,y_0,t)\\
  \tilde{y}(t) &=& \beta'(x_0,y_0,t)\\
  \lambda(t) &=& \gamma'(x_0,y_0,t)
\end{eqnarray}
where $\alpha'$,  $\beta'$, and $\gamma'$ are  suitable functions, and
$t$ is  an arbitrary parameter \citep[e.g.][]{pr17}.   Determining the
position  $(\tilde{x},\tilde{y})$ corresponding  to a  specific source
element position  $(x_0,y_0)$ and  wavelength $\lambda$  would involve
inverting the  equation for $\lambda(t)$  and using that value  in the
equations  for  $(\tilde{x}(t),\tilde{y}(t))$.   If the  form  of  the
equation permits, then wavelength can in principle be directly used as
the   parameter.    \LINEAR\   can   handle  the   general   case   of
\Eqn{eqn:genrep}, but in  the current code base  the simple polynomial
representation (\Eqn{eqn:trace} and \Eqn{eqn:disp}) is used.

The detector  pixels will have a  unique sensitivity as a  function of
wavelength,  which  in  standard   direct  imaging  is  calibrated  by
flat-field  images.  However  for grism  observations, the  problem is
significantly more complex as the wavelength of incident light depends
on the position of the source  on the detector.  Moreover the blending
of light from distinct regions leads to a convolution over wavelength,
and so  the flat-field  correction must  be taken  to be  position and
wavelength  dependent.  Additionally,  the average  pixel response  is
given by a  transmission curve, that transforms  the calibrated counts
to physical  units (in  the case  of HST  data from  e$^-$~s$^{-1}$ to
erg~s$^{-1}$~cm$^{-2}$~\AA$^{-1}$).  We  denote this  average response
as $S(\lambda)$ and show five detectable  orders for the G102 grism on
WFC3/IR in  \Fig{fig:sens}.  The  pixel-to-pixel deviations  from this
average  response  are included  in  the  form of  a  \emph{flat-field
  ``cube''}, which is typically given  as a polynomial over wavelength
whose coefficients encapsulate the spatial variations:
\begin{equation}\label{eqn:flat}
{\cal F}(x,y;\lambda)=\sum_{i=0}^{n}{\cal F}_i(x,y)\left(\frac{\lambda-\lambda_0}{\lambda_1-\lambda_0}\right)^i,
\end{equation}
where   $n$   is    the   order   of   the    flat-field   cube,   and
$(\lambda_0,\lambda_1)$ are  defined in  the calibration  process, but
are  roughly  the  wavelength  coverage of  the  grism  element.   The
coefficients $\{{\cal F}_i(x,y)\}$ are  images determined by comparing
monochromatic  flat-field  observations  \citep{kunt}.   We  implement
three   options  for   the   flat-field  cube:   unity  flat   (${\cal
  F}(x,y;\lambda)\!=\!1$), standard cube  described by \Eqn{eqn:flat},
and a single  direct-image flat-field \citep[\eg][]{brammer12}.  These
specifications   for   ${\cal   F}_i(x,y)$  are   suitable   for   the
reconstruction  of  WFC3/IR  grism   data,  but  \LINEAR\  can  handle
essentially arbitrary  forms for the wavelength-dependent  flat field,
without significant impacts on the processing.

There are geometric distortions in the  detector due to choices in the
optical design and imperfections in  the manufacture, which causes the
effective area of the pixels to vary across the detector.  Indeed, for
WFC3 much of the geometric distortion is by design, as a result of the
desire  to minimize  the number  of reflections  while correcting  the
spherical  aberration  and  maintaining   a  flat  focal  plane.   The
pixel-area maps  ${\cal P}(x,y)$ give  the area relative to  a nominal
pixel.  For WFC3/IR, the pixel  areas change by $\sim\!8\%$ across the
detector \citep[\eg][]{kali10}.

\section{Algorithm}\label{sec:algorithm}

\subsection{\LINEAR} \label{sec:code}

Inherent to our  working paradigm is that there  exists direct imaging
that  satisfies a  few requirements:
\begin{enumerate}
  \item covers $\sim\!2\times$  the area of the  instrument to account
    for sources  that disperse onto the  detector \citep[see Section~4
      of][]{pirzkal_isr},
  \item  considerably deeper  than  any individual  grism exposure  to
    minimize double-counting photometric noise  and ensure all sources
    are accounted for, and
  \item is at roughly the same  wavelength as sampled by the dispersed
    data   to  minimize   the  effect   of  any   wavelength-dependent
    morphological  effects and  changes in  the point-spread  function
    (PSF).
\end{enumerate}
We implement two  options to define the extraction  apertures for each
source, which we define as the collection of pixels that have a common
spectral shape.  The first method is to use a classic segmentation map
\citep[\eg][]{sex} with the same  world-coordinate system (WCS) as the
aforementioned direct imaging.  This approach is easy to implement and
familiar to many  users, but leads to the discrete  assignment of each
pixel to a  source (or the sky).  Therefore, our  second method allows
for the layering  of flux densities, associated  with multiple sources
that potentially overlap in the  direct image.  This is facilitated by
supplying  two multi-extension  \texttt{FITS} files  (MEF); the  first
expresses  the source  brightness  and second  defines the  extraction
pixels  (as a  binary  image).   Each extension  of  these files  must
contain  an  accurate WCS  and  represents  the information  for  each
source, therefore the  two images have the same  number of extensions.
This layered approach  additionally facilitates reconstructing spectra
for  overlapping regions  (such as  supernova/host galaxy,  bulge/disk
separation, or  overlapping galaxies).   Using the  MEF data,  we also
permit the reconstruction parameters (such as spectral sampling) to be
different   for  each   source.   Both   implementations  can   handle
disconnected  or \emph{island}  regions and  provide the  direct image
pixel-by-pixel  brightness,  which  is   used  in  the  reconstruction
process.

\onecolumngrid
\makeatletter
\global\@two@colfalse
\makeatother

\begin{algorithm}[H]
 \caption{Computing $W$-matrix Element}\label{alg:matrix}
 \begin{algorithmic}[1]
   \State \Input\ list of dispersed images, direct image, segmentation map, and wavelength resolution
   \For{\each\ dispersed image {\bf in} image list}
     \For{\each\ object {\bf in} segmentation image}
       \For{\each\ pixel {\bf in} the object}
         \State transform the pixel corners to dispersed image using WCSs (including distortion)
         \For{\each\ wavelength associated {\bf with} the object}
           \State invert dispersion (\Eqn{eqn:disp}) to determine path length for given wavelength
           \State invert path length (\Eqn{eqn:path}) to compute $\tilde{x}$-coordinate
           \State compute $\tilde{y}(\tilde{x})$ from spectral trace (\Eqn{eqn:trace})
           \State convert trace to dispersed-image coordinates (\Eqn{eqn:offset})
           \State compute overlapping area ($a(\lambda)$) with dispersed image pixel grid

           \State compute weight (\Eqn{eqn:weight})           
         \EndFor
       \EndFor
       \State sum weights over unique combinations of $(x,y,\lambda)$
       \State record the {\it object-dispersion table} (ODT) of $(x,y,\lambda,W)$
     \EndFor
   \EndFor 
   \State load the ODTs and sum $w$ in wavelength bins --- the bin widths and locations can vary between objects
   \State divide weights by image uncertainty for least-squares minimization (see \Eqn{eqn:uncertainty})
   \State sum over unique combinations of $(x,y,\lambda)$ --- now have multiple objects to sum
   \State \Output\ sparse matrix stored in COO-notation ($\vartheta,\varphi,W_{\vartheta,\varphi}$)
 \end{algorithmic}
\end{algorithm}
\makeatletter
\@two@coltrue
\makeatother
\twocolumngrid

The brightness detected in a dispersed-image pixel is the sum
over all sources and wavelengths:
\begin{equation}\label{eqn:sum}
  {\cal G}_{x,y,i} = \sum_\lambda \sum_j^{N_{\rm obj}} W_{x,y,i,\lambda,j}\,f_{\lambda,j},
\end{equation}
where ${\cal G}_{x,y,i}$ is the  measured brightness in the \ith-grism
image  at   pixel  $(x,y)$,   $f_{\lambda,j}$  is  spectrum   for  the
\jth-object,  and  $W$  is  a  collection  of  transformation  factors
describing the dispersed image signal  for a unit spectrum.  The value
of $W$ includes the flux distribution of the source, and is explicitly
summed over all direct-image pixels  in the source. For typical sparse
fields, many of  the $W$-elements will be zero  by construction, since
only a  few sources can  contribute flux to a  given pixel at  a given
wavelength.  By grouping the measurement (``the knowns'') and spectral
(``the unknowns'') indices respectively:
\begin{eqnarray}
  (x,y,i) &\rightarrow& \vartheta\\
  (\lambda,j) &\rightarrow& \varphi,
\end{eqnarray}
\Eqn{eqn:sum} becomes a simple matrix product:
\begin{equation}\label{eqn:mat}
  {\cal G}_{\vartheta} = \sum_{\varphi} W_{\vartheta,\varphi}\,f_{\varphi}.
\end{equation}
\LINEAR\ populates the elements of the $W$-matrix by iterating through
the canon of dispersed images and objects in the extraction apertures.
Since  the direct  image and  segmentation map  may have  an arbitrary
pixelation, it often  necessary to subsample the  wavelength to ensure
that  each  dispersed  image  locations  represents,  with  sufficient
fidelity, the  spectral energy distribution  (SED) of each  source. We
implement the subsampling  as an integer number of  steps smaller than
the              native              spectral              resolution:
$\delta\lambda\!\approx\!N^{-1}\,\mathrm{d}\lambda/\mathrm{d}s$, where
$\mathrm{d}\lambda/\mathrm{d}s$ is  the field-averaged  dispersion for
the        detector         (for        WFC3/G102         it        is
$\mathrm{d}\lambda/\mathrm{d}s\!\approx\!25$~\AA~pix$^{-1}$)  and  $N$
is  the  subsampling factor.   We  explored  various choices  for  the
subsampling  factor, but  found that  $N\!=\!5$ is  a good  compromise
between computational  constraints and spectral fidelity  for a direct
image with  pixels $\sim\!16\times$  smaller than the  native detector
(for WFC3/IR the native pixel scale is $\sim\!0\farcs12$).  However we
caution that the best value of $N$ likely varies on the properties of the
direct imaging in question.


In  this formalism,  $W$  is  a linear  operator  that transforms  the
spectra  $f_\varphi$  into  the  observed  pixel  brightnesses  ${\cal
  G}_\vartheta$.  The bulk of the processing for \LINEAR\ is computing
the   $W$-matrix  elements;   we  summarize   these  calculations   in
Algorithm~(\ref{alg:matrix}).
The matrix elements  (weights in the linear equations) are  given as a
product of the instrumental effects described above:
\begin{equation}\label{eqn:weight}
  W = a(\lambda)\,{\cal I}(x_0,y_0)\,{\cal F}(x,y;\lambda)\,{\cal P}(x,y)\,S(\lambda)\,\delta\lambda
\end{equation}
where $a(\lambda)$ is the relative pixel area from the reprojection at
some  wavelength, ${\cal  I}(x_0,y_0)$ is  the \emph{normalized} image
brightness from the direct  image, ${\cal F}(x,y;\lambda)$ is the flat
field,  ${\cal P}(x,y)$  is the  correction from  the  pixel-area map,
$S(\lambda)$ is the average  sensitivity curve, and $\delta\lambda$ is
the   subsampled   bandwidth.


The optimal  set of object  spectra will minimize  the goodness-of-fit
metric derived from maximizing a Gaussian likelihood:
\begin{equation}\label{eqn:chi2}
  \chi^2 = \sum_{\vartheta}\left(\frac{\sum_{\varphi}W_{\vartheta,\varphi}f_{\varphi}-{\cal G}_{\vartheta}}{{\cal U}_{\vartheta}}\right)^2
\end{equation}
where ${\cal  U}_{\vartheta}$ is  the estimated uncertainty  for datum
$\vartheta$.  These uncertainties can  be factored into the $W$-matrix
and the pixel-by-pixel flux measurements (${\cal G}_{\vartheta}$) as:
\begin{eqnarray} \label{eqn:uncertainty}
  W_{\vartheta,\varphi}&\rightarrow&W_{\vartheta,\varphi}/{\cal U}_{\vartheta}\\
  {\cal G}_{\vartheta}&\rightarrow&{\cal G}_{\vartheta}/{\cal U}_{\vartheta}.
\end{eqnarray}
\EQN{eqn:chi2}  can  be  extended  to  a  damped-least  squares  in  a
simplified-matrix notation of:
\begin{equation}\label{eqn:lsq}
  \hat{f}=\underset{f}{\argmin} \left(||W\,f - {\cal G}||^2+\ell\,||W||_F^2\,||f||^2\right),
\end{equation}
where $\hat{f}$ is the optimized  solution, $||W||_F$ is the Frobenius
norm of $W$, and $\ell$ is a damping parameter that imposes smoothness
in the final  solution.  We defer discussion of  the damping parameter
and its  effects to \Sect{sec:damp} and  \Sect{sec:hudf}.  Unlike most
presentations, we  include the  Frobenius norm in  the second  term so
that the damping parameter is dimensionless.

The size of the $W$-matrix will be  the number of knowns by the number
of  unknowns.  The  number of  knowns is  the total  number of  pixels
analyzed, which includes all the pixels that contain source flux (that
may be as large as the number of images times the number of pixels per
image, but is generally less).  The  number of unknowns is roughly the
number of objects times the number  of flux elements per object (to be
clear, \LINEAR\ allows the number of flux elements to be different for
each object).   For a  deep dataset with  WFC3/IR \citep[\eg][]{figs},
there  may be  $\sim\!100$~images with  $\sim\!10^6$~pixels each  (but
only $\sim\!10\%$  of these  pixels may have  source flux)  to extract
$\sim\!1000$~objects for $\sim\!100$~spectral elements.  Therefore the
$W$-matrix will  have dimensionality of $10^7\times10^5$,  however the
vast majority ($\gtrsim\!99.9$\%) of  these elements are exactly zero,
since any  one source will affect  only a small fraction  of pixels on
any  one  dispersed image  at  any  wavelength.  Therefore  we  employ
several sparse-matrix techniques,  specifically storing the $W$-matrix
in  the  coordinate-list  (COO)  format and  use  the  LSQR  algorithm
\citep{ps82}.

The  LSQR  algorithm  estimates  the  uncertainty  on  the  parameters
(\ie\ the  reconstructed spectra in  this case), however  these values
will  be underestimated  for $\ell\!\neq\!0$.   Paradoxically, a  very
large damping parameter implies  vanishingly small uncertainty, as the
optimization  depends  only  on  the  variance  of  the  reconstructed
spectra; so that \Eqn{eqn:lsq} becomes:
\begin{equation}\label{eqn:largeell}
  \hat{f}=\underset{f}{\argmin} \left(||f||^2\right)
\end{equation}
when  $\ell\!\rightarrow\!\infty$.   Therefore, the  optimal  solution
will be  when all elements  of $f$ are  exactly zero.  To  account for
this, we  implement a  Markov Chain Monte  Carlo (MCMC)  simulation to
sample from  the likelihood, which is  given by $\ln(L)\!=\!-\chi^2/2$
(see \Eqn{eqn:chi2} for  the definition of $\chi^2$).   In many cases,
this will be of exceedingly  high dimensionality, therefore we compute
the uncertainties for each source  sequentially by holding the spectra
of  the  remaining  sources  fixed  to the  values  found  by  solving
\Eqn{eqn:lsq}  with  LSQR.   Our sequential  uncertainty  analysis  is
formally correct for the limiting  case of no overlapping dispersions,
but does not include any  correlations induced by overlapping spectra.
We  have  verified the  uncertainty  estimates  coming from  the  MCMC
sampler are consistent with the  uncertainties reported by other means
\citep[\eg][]{figs}.   We further  discuss additional  details of  the
damping parameter in \Sect{sec:damp}.

The      majority     of      the     code      is     written      in
\texttt{IDL}\footnote{\href{https://www.harrisgeospatial.com/docs/using_idl_home.html}{https://www.harrisgeospatial.com/docs/using\_idl\_home.html}}
with the  exception of two  key steps:  (1) the area  computations are
performed via  the Sutherland-Hodgman  algorithm with  \texttt{C} code
provided      by     J.~D.~Smith      \citep[as     developed      for
  \texttt{CUBISM}:][]{polyclip},  and  (2)  we  translated  the  scipy
implementation of  the LSQR algorithm into  \texttt{C}.  However, both
components  can default  to \texttt{IDL}-only  implementations if  the
\texttt{C} code is not successfully  compiled.  We have integrated the
multithreading package  \texttt{OpenMP} into  the components  that are
written  in   \texttt{C},  but   provide  non-threaded   versions  for
simplicity.

\subsection{Bad Pixels and Additional Spectral Orders}

It is important to note that  we remove all pixels from $\{{\cal G}\}$
that are flagged  in the bad-pixel masks (BPXs) or  having flux coming
from  additional  spectral  orders\footnote{We   expect  users  to  be
  primarily focused  on the $+1$~order  and consider all others  to be
  contaminating, however it is possible to modify this behavior.}.  To
identify  pixels from  additional  orders, we  follow  a very  similar
procedure to that described  in Algorithm~\ref{alg:matrix}, however to
expedite the calculations  we do not consider each  pixel belonging to
an object, but rather group the pixels into a convex hull.  This would
amount  to   a  considerable  loss   in  spectral  fidelity   if  this
prescription were used for building  the $W$-matrix.

\subsection{Additional Capabilities}

The vast majority of the processing efforts for \LINEAR\ deal with 
populating the $W$-matrix elements.  However once $W$ is fully formed, 
there are many useful calculations that can be readily performed:
\begin{itemize}
  \item predicting contamination estimates --- valuable for planning 
    observations of a select number of high-value science targets;
  \item providing average wavelength for sources; and 
  \item excising two-dimensional grism images for sources.
\end{itemize}
As these  are not  the primary  goal of this  work, we  defer detailed
discussion  of  their  usability  to  the  \LINEAR\  reference  manual
provided with the codebase \citep{manual}.

\subsection{Preprocessing}

\LINEAR\  simply   represents  a   new  paradigm   for  reconstructing
one-dimensional spectra from a canon  of dispersed images, and several
pre-processing steps must be performed.

\subsubsection{Background Subtraction} \label{sec:backsubtract}

The sky background in a grism  image is far more complex than standard
direct imaging,  and is typically  estimated by fitting a  \emph{master
  sky}  image  to the  sky  pixels  in  an image  \citep[as  discussed
  in][]{grapes}. This inherently assumes the  sky flux is dominated by
a  single spectral  component,  whereas \citet{brammer14}  demonstrate
that for WFC3/IR there are  at least two distinct spectra contributing
to the  sky background  (\eg~zodical light and  \ion{He}{1} emission).
Therefore  the  fitting  process  becomes quite  a  bit  more  complex
\citep[see the  \S~6.~Appendix:~Iterative Inversion of][]{brammer_isr}.
This  becomes  even  more  problematic  should  one  of  the  spectral
components vary with  time, and the standard  ramp-fitting in infrared
detectors   must   be   incorporated  in   sky   background   modeling
\citep[\eg][]{pirzkal17}.   Since it  is impossible  to predict  which
algorithm (single  master sky, multiple sky  components, time-variable
components) a dataset might require, we instead assume the images have
been background  subtracted by some means.   In \Sect{sec:backsim}, we
show how residual levels of flux  will adversely affect the quality of
the extracted spectra.

\subsubsection{Astrometric Registration}

As described  in \Sect{sec:primer},  HST grism  observations typically
include  obtaining a  shallow  pre-image to  accompany each  dispersed
image (or  set of dispersed images  with known dithers) to  refine the
astrometry of the dispersed data.   There are numerous tools available
for such processing \citep[such as][]{gonzaga,avila}, and so we do not
attempt any astrometric corrections.  The possible impact of registration
errors is discussed in \Sect{sec:astro_error}.

If the  direct and  dispersed images  are not taken  at the  same time
\citep[as  may  be  the  case for  the  WFIRST  High-Latitude  Survey:
  HLS;][]{wfirst}, then appropriate preprocessing must be carried out
to ensure that the direct  and dispersed images are correctly aligned.
Some concepts and methods to this end have been discussed or developed
by \citet{dixon} and \citet{bohlin15}.

\section{Tests of \LINEAR} \label{sec:tests}

\begin{figure}
  \includegraphics[width=0.47\textwidth]{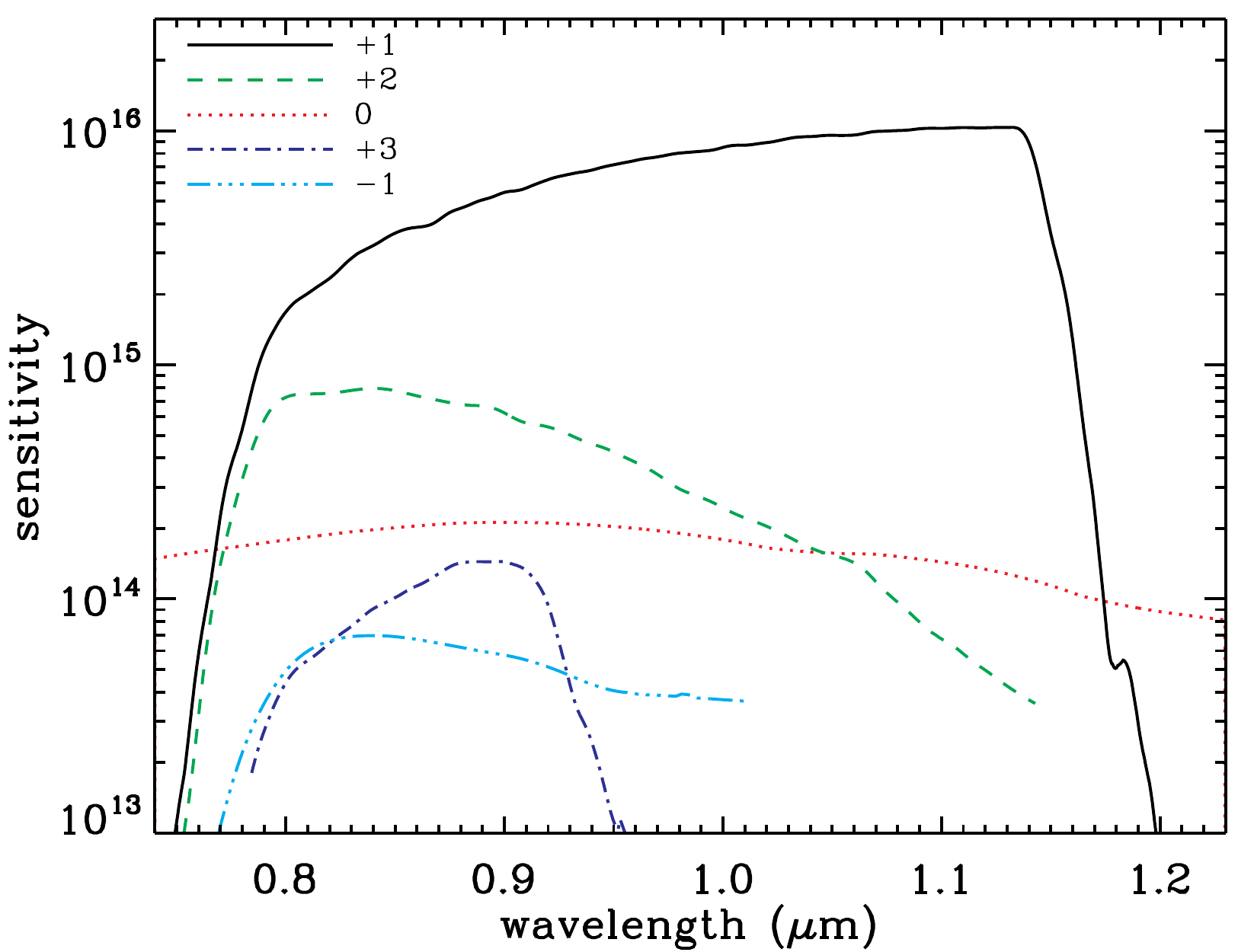}
  \caption{The  sensitivity  curves  for  HST WFC3/G102  in  units  of
    e$^-$~s$^{-1}$  per  erg~s$^{-1}$~cm$^{-2}$~\AA$^{-1}$. Each  line
    represents a different  spectral order, as described  in the upper
    left:  $+1^\mathrm{st}$  (solid black),  $+2^\mathrm{nd}$  (dashed
    green), $0^\mathrm{th}$ (dotted red), $+3^\mathrm{rd}$ (dot-dashed
    purple),       and      $-1^\mathrm{st}$       (dot-dot-dot-dashed
    cyan).\label{fig:sens}}
\end{figure}

For the following tests of  \LINEAR, we consider observations from the
\emph{Hubble Space Telescope}  using the Wide-Field Camera  3 with the
G102 grism element with pre-imaging  through the F105W filter. For all
of the  tests described below,  we use \LINEAR\ to  generate simulated
grism images  through the  G102/F105W combination, which  includes the
best  available estimates  of  the spectral  trace, dispersion,  field
edges, average  sensitivity, wavelength-dependent flat field,  and the
pixel-area map.  By accounting for filter-wedge offsets \citep{sabbi},
it   is  possible   to   consider   other  grism/filter   combinations
\citep{pirzkal_isr}.   In \Fig{fig:sens},  we show  the field-averaged
sensitivity of the G102 element to set the stage.

\subsection{Uncertainties and Error Propagation}

Below  we describe  a series  of Monte  Carlo simulations  designed to
assess  the impact  of the  uncertainties in  the calibrations  on the
final spectroscopy.  In each simulation, we consider a single Gaussian
point source with a flat spectrum in $f_{\lambda}$.

\subsubsection{Astrometric Registration} \label{sec:astro_error}

\begin{figure}
  \includegraphics[width=0.47\textwidth]{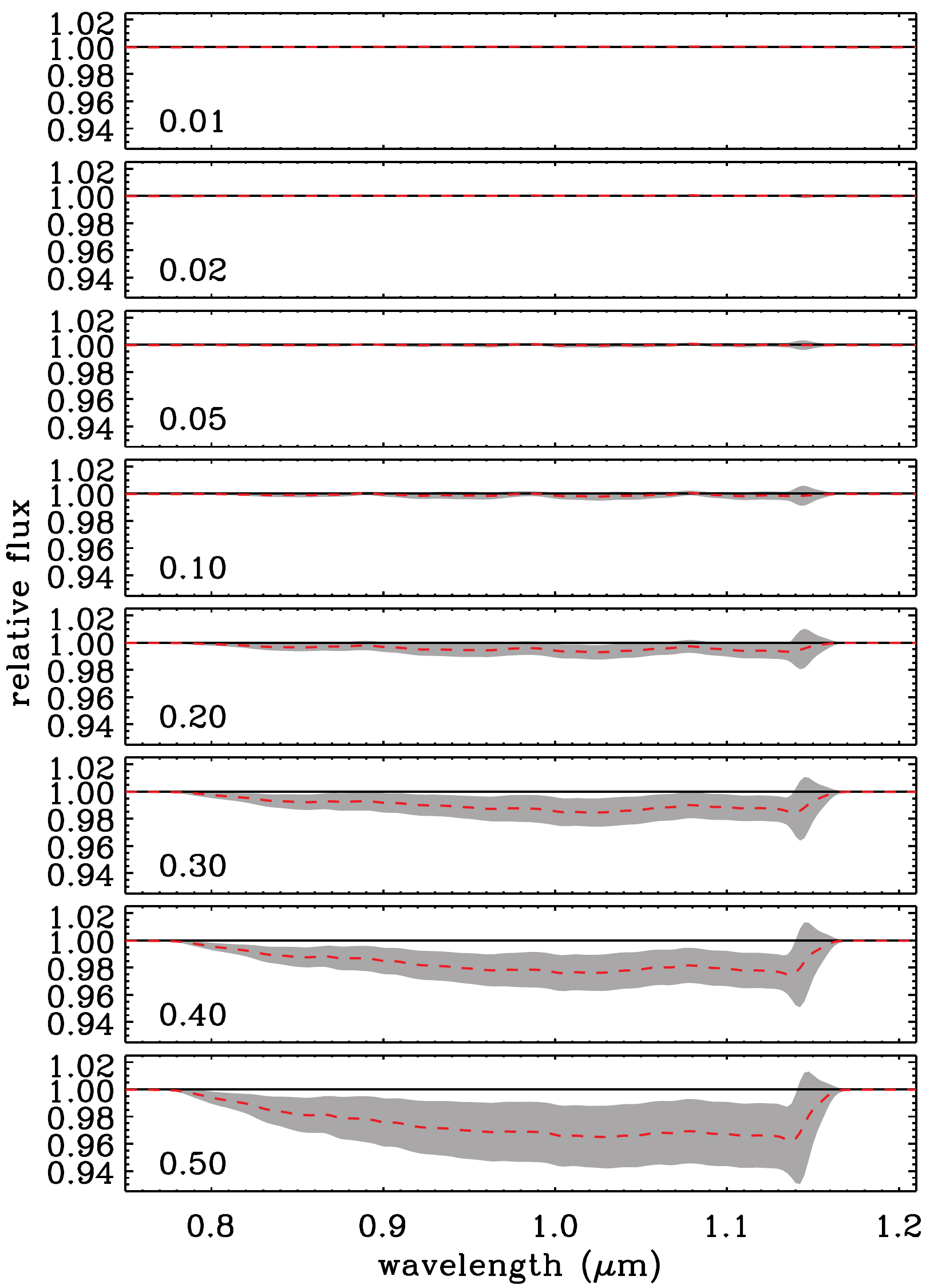}
  \caption{The   relation   between   astrometric   registration   and
    spectrophotometric uncertainty.  In the  upper left corner of each
    panel, we give  the uncertainty in the  \texttt{CRVAL} keywords in
    units of  native pixel scale ($\sim\!0\farcs12$).   For this Monte
    Carlo  simulation  (described   in  \Sect{sec:astro_error}),  we
    consider a  single Gaussian point  source with a flat  spectrum in
    $f_{\lambda}$  (black   line)  observed   at  four   orients.   By
    simulating this configuration many times with Gaussian noise added
    to the astrometry, we synthesize  the average spectrum (red dashed
    line)  and  standard deviation  (gray  region)  as a  function  of
    wavelength.  For typical HST/WFC3  observations, the astrometry is
    easily refined  to $\sim\!0.1$~pix  and therefore \LINEAR\  is not
    expected   to   introduce  any   significant   ($\lesssim\!0.1\%$)
    additional spectrophotometric noise.  \label{fig:astro}}
\end{figure}

Since  we are  extracting the  spectra on  a native  pixel scale,  the
precise registration of the dispersed  image to the existing images is
crucial.   We simulate  a Gaussian  object for  four distinct  orients
$\{0^\circ,90^\circ,180^\circ,270^\circ\}$  with no  photometric noise
(as  the goal  here  is  to isolate  the  effects  of astrometry,  and
photometric noise will be considered  below). We add a Gaussian random
offset  to the  \texttt{CRVAL} keywords  to simulate  an error  in the
astrometric registration, and reconstruct the one-dimensional spectrum
with the \LINEAR\ algorithm.  We  repeat this procedure many times and
show the average (red dashed)  and standard deviation (gray region) as
a function  of wavelength in  \Fig{fig:astro}.  The input  spectrum is
shown in  black and the standard  deviation of the Gaussian  offset in
native pixels is given in the lower left corner of each panel.

For  most  realistic  situations,  it is  fairly  straight-forward  to
astrometrically align  images to  the $\sim\!0.1$~pix  in the  case of
WFC3/IR.   Based  on our  Monte  Carlo  simulation, we  estimate  this
astrometric uncertainty  introduces an  error of  $\lesssim\!1$\%.  We
also   find   that  for   the   wavelengths   with  high   sensitivity
($0.8\!\lesssim\!\lambda\!\lesssim\!1.1~\mu$m),       the      typical
photometric             uncertainty            scales             like
$\sigma_f\!\approx\!0.03\,\sigma_r$, where $\sigma_r$ is the offset in
the  astrometry  in native  pixels  (\ie\  the  values listed  in  the
lower-left corner of each panel  in \Fig{fig:astro}) and $\sigma_f$ is
uncertainty on the spectrum for a source brightness of $Y\!=\!18$~mag.

\subsubsection{Background Subtraction}\label{sec:backsim}

\begin{figure}
  \includegraphics[width=0.47\textwidth]{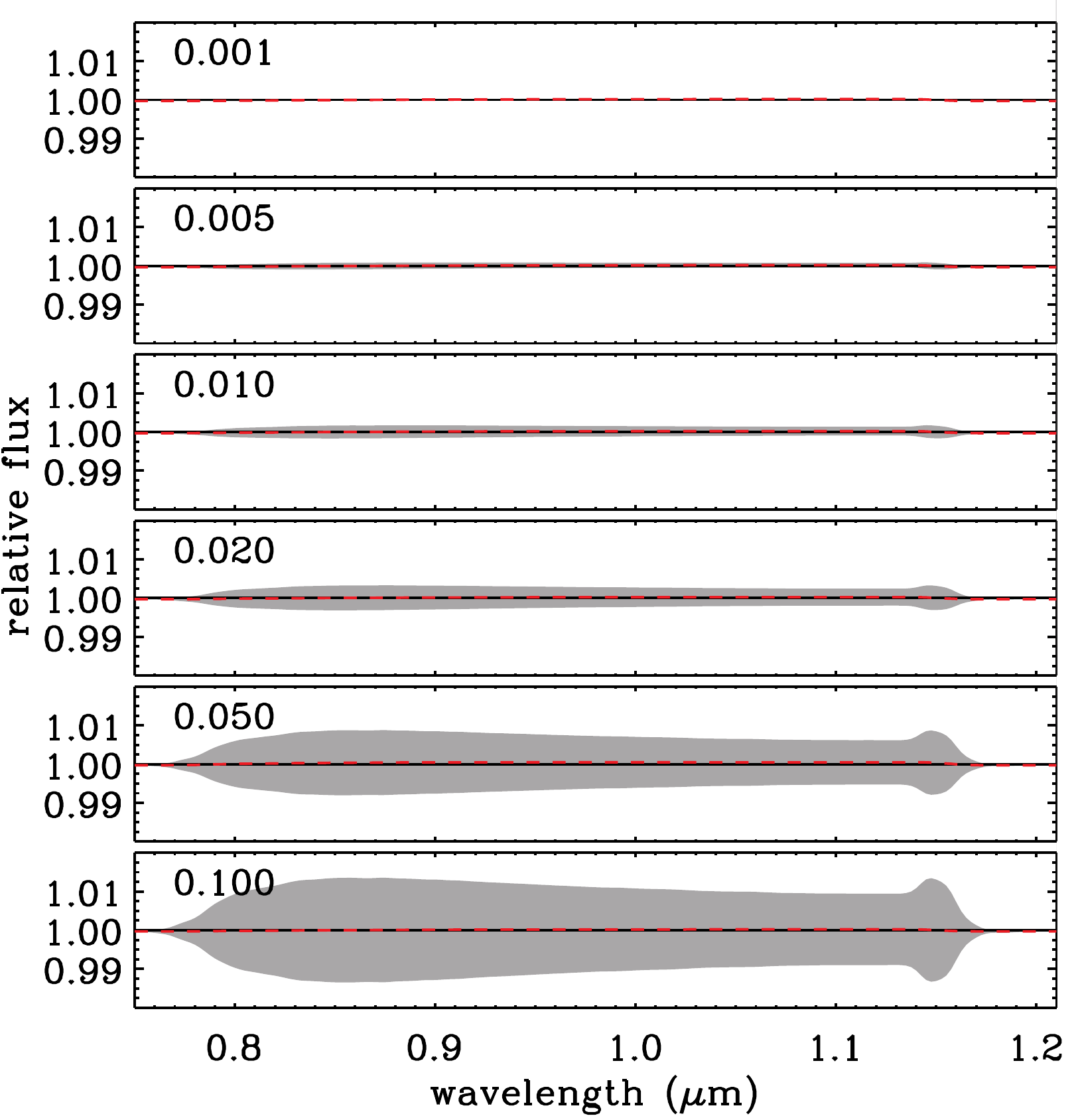}
  \caption{The  effect  of  residual   background  level  on  spectral
    uncertainty.  In the upper left corner  of each panel, we give the
    uncertainty  for  the  background  level  in  $e^-$~s$^{-1}$.   We
    describe this simulation in  \Sect{sec:backsim}.  Here the colors,
    plot symbols, and axes are  the same as in \Fig{fig:astro}.  Based
    on  a collection  of  WFC3/IR data,  \citet{pirzkal17} estimate  a
    typical residual  level of  $\sim\!0.01$~$e^-$~s$^{-1}$, therefore
    the      typical     induced      spectral     uncertainty      is
    $\sim\!0.1\%$. \label{fig:back}}
\end{figure}

We now consider the Gaussian source through the same four orients, but
impose  a  Gaussian random  pedestal  sky  brightness.  For  each  sky
background  uncertainty,  we generate  100  realizations  of the  four
orients.  In \Fig{fig:back}, we  show the averaged extracted spectrum,
and  the  colors  and  plot  symbols  have  the  same  meaning  as  in
\Fig{fig:astro}.   In   a  recent  Instrument-Science   Report  (ISR),
\citet{pirzkal17} demonstrate  that one can remove  the sky background
levels  to   $\sim\!0.01$~e$^-$~s$^{-1}$  using  a   constant  zodical
background and  time-varying helium emission  model \citep{brammer14}.
Following that prescription, we  estimate the inaccuracies introduced
from improper background subtraction to be $\sim\!0.1\%$. We find that
the  photometric   uncertainty  scales   roughly  as   the  background
subtraction uncertainty: $\sigma_f\!\approx\!0.13\,\sigma_B$.

It  may  be possible  to  further  reduce the  background  subtraction
accuracy    to   $\sim\!0.001$~e$^-$~s$^{-1}$    with   a    smoothed,
column-averaged       correction        and/or       global       mean
\citep[\eg][]{figs}. Since the typical  background accuracy is already
a sub-dominant  term in the  uncertainty budget,  we do not  carry out
further   corrections.   In   \Sect{sec:mods},  we   discuss  possible
modifications to  our algorithm that  may provide such  refinements in
the background.

\subsubsection{Photometry and Signal-to-Noise}\label{sec:photom}

\begin{figure}
  \includegraphics[width=0.47\textwidth]{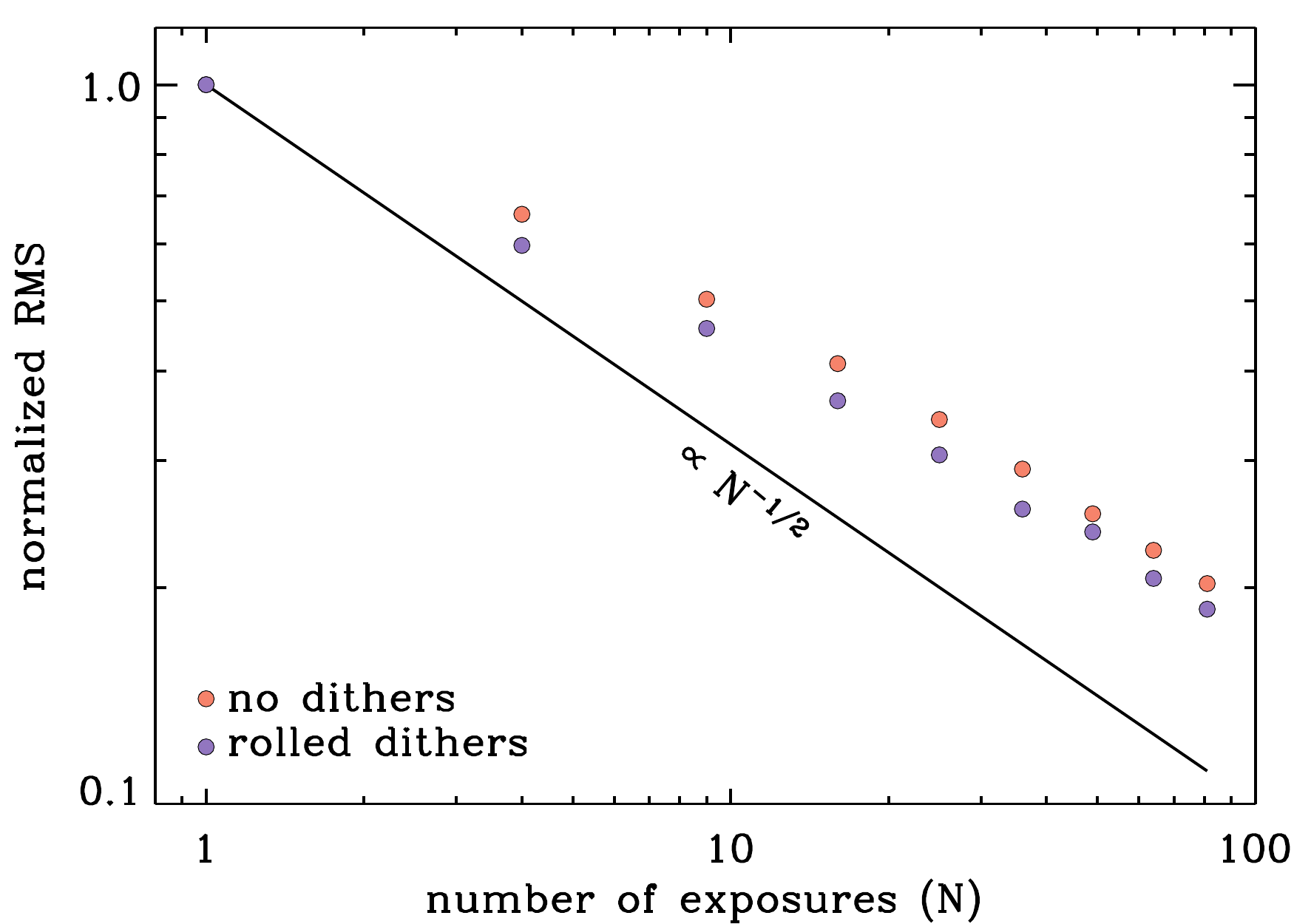}
  \caption{Spectrophotometric    uncertainty.     As   described    in
    \Sect{sec:photom},  we  consider  a   series  of  $N$  independent
    exposures, each  with noise  consistent with  a $1200$~s  image in
    WFC3/G102.  We then extract the  source in various combinations of
    these   images  to   show  the   decrease  in   spectrophotometric
    uncertainty with exposure time.  The  red and blue points show the
    uncertainty for undithered and roll-dithered images, respectively.
    The  solid   black  line   represents  the  standard   scaling  of
    $\propto\!N^{-1/2}$,  and each  set  of points  are normalized  to
    unity  at  $N\!=\!1$.  The  simulated  data  scale more  shallowly
    ($\propto\!N^{-0.4}$).}\label{fig:photom}
\end{figure}

To assess the  relationship between total exposure time  (or number of
images) and inferred signal-to-noise, we consider the same source with
flat  spectrum in  $f_\lambda$ as  described above.   However here  we
explicitly include Gaussian random  noise consistent with a $1200$~s
exposure with G102 on WFC3/IR.  We  create $N$ images of this scenario
and extract  the spectrum, then  repeat for 100~realizations  of these
images to estimate the variance for a fixed number of input images. We
consider  two  cases  for  the  orientation of  the  $N$  images:  (1)
perfectly  coaligned,  with  no  rotation;  and  (2)  rotated  by  $N$
different angles,  uniformly distributed in the  range $[0,2\pi]$.  We
repeat this  procedure for  different values of  the number  of images
$N$,  ranging from  1 to  100.  \Fig{fig:photom}  shows the  resulting
spectrophotometric  uncertainty,  expressed  as the  root-mean  square
(RMS), averaged in bins of wavelength, as a function of $N$, scaled to
the  single-image   value.   We  find   that  the  RMS   error  scales
approximately as  $N^{-0.4}$ for both  the case with no  rotation (red
points) and with uniformly distributed rotations (blue points).

\subsection{The Effect of Damping} \label{sec:damp}

\begin{figure}
  \includegraphics[width=0.47\textwidth]{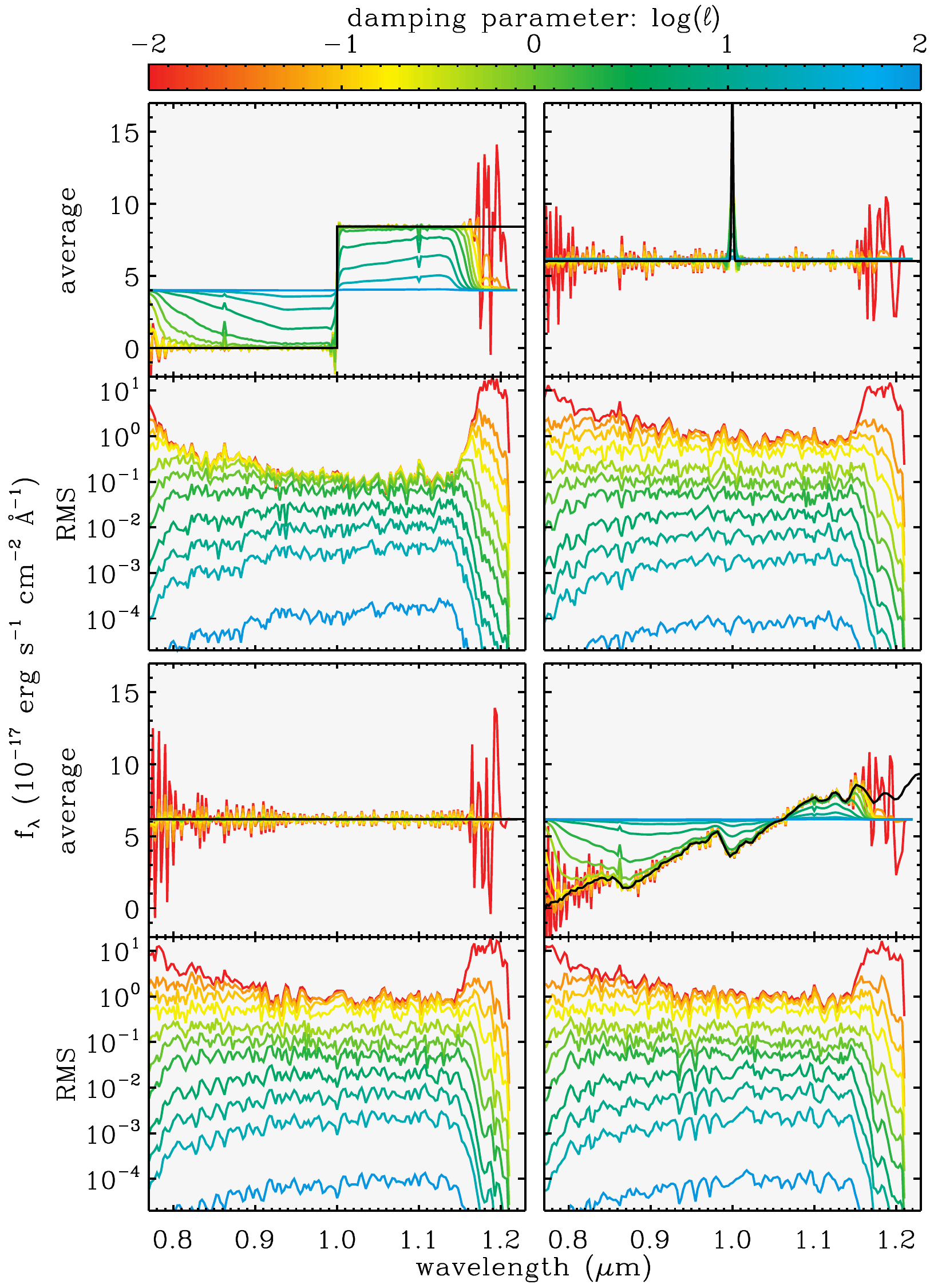}
  \caption{Effect of  the damping parameter on  the extracted spectrum
    of a  Gaussian source.  The four  panels are split into  a top and
    bottom  sub-panel   that  show  the  average   and  RMS  spectrum,
    respectively, for different input  spectra, represented by a black
    line (top left:  step function, top right:  constant with emission
    line,  and bottom  right: L5V  brown dwarf).   These sources  were
    extracted  with  a  range   of  damping  parameters  ($\ell$,  see
    \Eqn{eqn:lsq}), as indicated with the color bar.}\label{fig:damp}
\end{figure}

The damping coefficient  $\ell$ in \Eqn{eqn:lsq} applies  a penalty to
solutions of $\{f\}$ that have high  variance, which has the effect of
damping high-frequency  oscillations.  In  the case  of \Eqn{eqn:lsq},
the  oscillations  are  damped  with respect  to  $f\!=\!0$  (for  all
elements of  $f$), but this  can be  extended to an  arbitrary damping
target ($f_0$) as
\begin{equation}\label{eqn:lsq_damp}
  \hat{f}'=\underset{f'}{\argmin}\left(||W\,f' - {\cal G'}||^2+\ell\,||W||_F^2\,||f'||^2\right),
\end{equation}
where   $f'\!=\!f-f_0$,    ${\cal   G'}\!=\!{\cal    G}+W\,f_0$,   and
$\hat{f}\!=\!\hat{f}'+f_0$.  To  visualize the  effect of  damping, we
consider  four SEDs  that  serve  to highlight  the  range of  effects
typical in  many grism  surveys: a constant  in $f_{\lambda}$,  an L5V
brown dwarf\footnote{From  the SpeX  Prism Library maintained  by Adam
  Burgasser                                                         at
  \href{http://www.browndwarfs.org/spexprism}{http://www.browndwarfs.org/spexprism}.},
a step function  with break at $\lambda_{\rm  obs}\!=\!1~\mu$m, and an
emission line source with  line at $\lambda_{\rm obs}\!=\!1~\mu$m (the
line has a Gaussian profile  and an observer-frame equivalent width of
$W_{\rm  obs}\!\approx\!-50$~\AA).   In  all  cases,  we  implement  a
Gaussian source profile normalized  to $Y\!=\!18$~mag that is observed
in 100~dithered images,  each having realistic noise.   We extract the
spectrum   for    each   SED    with   various    damping   parameters
$\ell\!\in\!\{0,0.01,0.05,0.1,0.2,0.5,1,2,5,10,20,100\}$, damped  to a
constant           $f_{\lambda}$           of           $Y\!=\!18$~mag
($f_{\lambda}\!\approx\!6.1\times10^{-17}$~erg~s$^{-1}$~cm$^{-2}$~\AA$^{-1}$),
and iterate this  procedure 100 times to estimate  the variance around
the mean.  In \Fig{fig:damp}, we  show the extracted spectra, averaged
over the  100 iterations,  with each  value of  $\ell$ indicated  by a
different color;  the black line  represents the input  spectrum.  The
red  lines show  the  spectrum extracted  with $\ell\!=\!0.01$,  which
clearly reproduce the  input (black line) spectra  the best.  However,
they  also show  the  largest  variance around  the  mean,  and so  we
consider these  the \emph{hot}  extractions.  On  the other  hand, the
blue lines ($\ell\!=\!100$) are highly  smoothed versions of the input
(\ie\  the   damping  target  $f_0$).   In   fact,  these  \emph{cold}
extractions are  so damped, they are  often flat spectra close  to the
average brightness.

This simulation raises a challenging question: \emph{How to select the
  ideal  value for  $\ell$?}   After all,  spectra reconstructed  with
little  damping  (small  $\ell$)  are  close  to  the  input  spectrum
(\ie\ the  damping target)  at the  cost of  increased noise  from the
inversion process.  Moreover,  LSQR may bias the  extracted toward the
damping target  for $\ell\!\gg\!0$;  see the  blue spectra  plotted in
\Fig{fig:damp}, particularly  the top left panel.   This effect arises
when the  damping target is a  poor match to the  underlying spectrum,
which  is  generally unknown  in  the  reconstruction phase.

\citet{hansen}  reviews  analysis of  the  {\it  L-curve}, a  plot  of
$\log||Wf-{\cal  G}||^2$ and  $\log||f||^2$ that  is parameterized  by
$\ell$, that offers a compromise  between fitting the observations and
damping fluctuations. The  point where the local  curvature is maximum
is  considered the  optimal  damping parameter  in reconstructing  the
incident  spectra. \citet{cult}  present  an  iterative algorithm  for
locating  this  critical  point  based on  the  golden-section  search
method.  At this  time, we do not implement this  approach; instead we
sweep through a range of the  damping parameter to locate the critical
point, as discussed in \Sect{sec:hudf}.

Following  the  preceding  discussion in  \Sect{sec:code},  the  LSQR
algorithm formally provides a mechanism  for computing the variance of
the solution.  Now for a  non-zero damping target  $f_0\!\neq\!0$, the
optimal solution is no longer  identically zero, but rather approaches
$f_0$.  The  concern that  LSQR significantly underestimates  the true
uncertainty is still present, and so the uncertainty derived from the
MCMC sampling is preferred.

\subsection{Multiple Sources and Overlapping Traces} \label{sec:multiple}

\begin{figure}
  \includegraphics[width=0.47\textwidth]{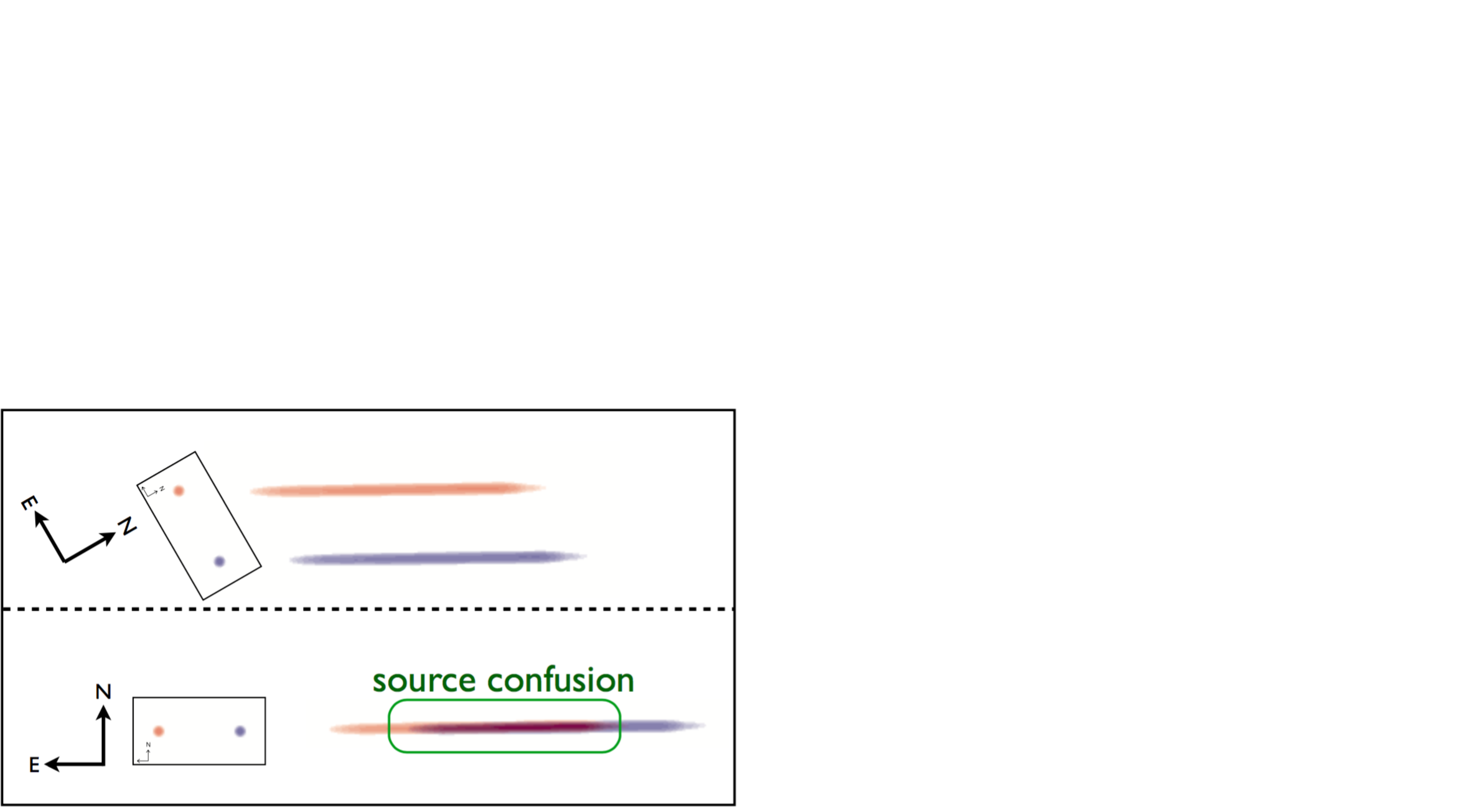}
  \caption{Illustrative example of spectral confusion for two sources.
    Here we consider two Gaussian sources  (shown in the insets as red
    and blue) whose relative position in the sky is fixed, but we view
    them under two different position angles (top and bottom figures).
    The $1^\mathrm{st}$  order traces of  each source is shown  in its
    corresponding color (the colors here  are chosen to highlight the
    overlap and do not refer to  the spectral shape of either source).
    In the  bottom panel,  the $1^\mathrm{st}$ order  dispersions from
    the  two sources  considerably  overlap, leading  to  a region  of
    spectral  confusion  (as  indicated  with  the  green  rectangle).
    However when  these objects are  observed at a  different position
    angle,  then  their  dispersions  do  not  overlap,  yielding  two
    unconfused  $1^\mathrm{st}$ order  traces.  This  example provides
    the  main motivation  behind  the development  of this  algorithm.
  }\label{fig:confusion}
\end{figure}

\begin{figure}
  \includegraphics[width=0.47\textwidth]{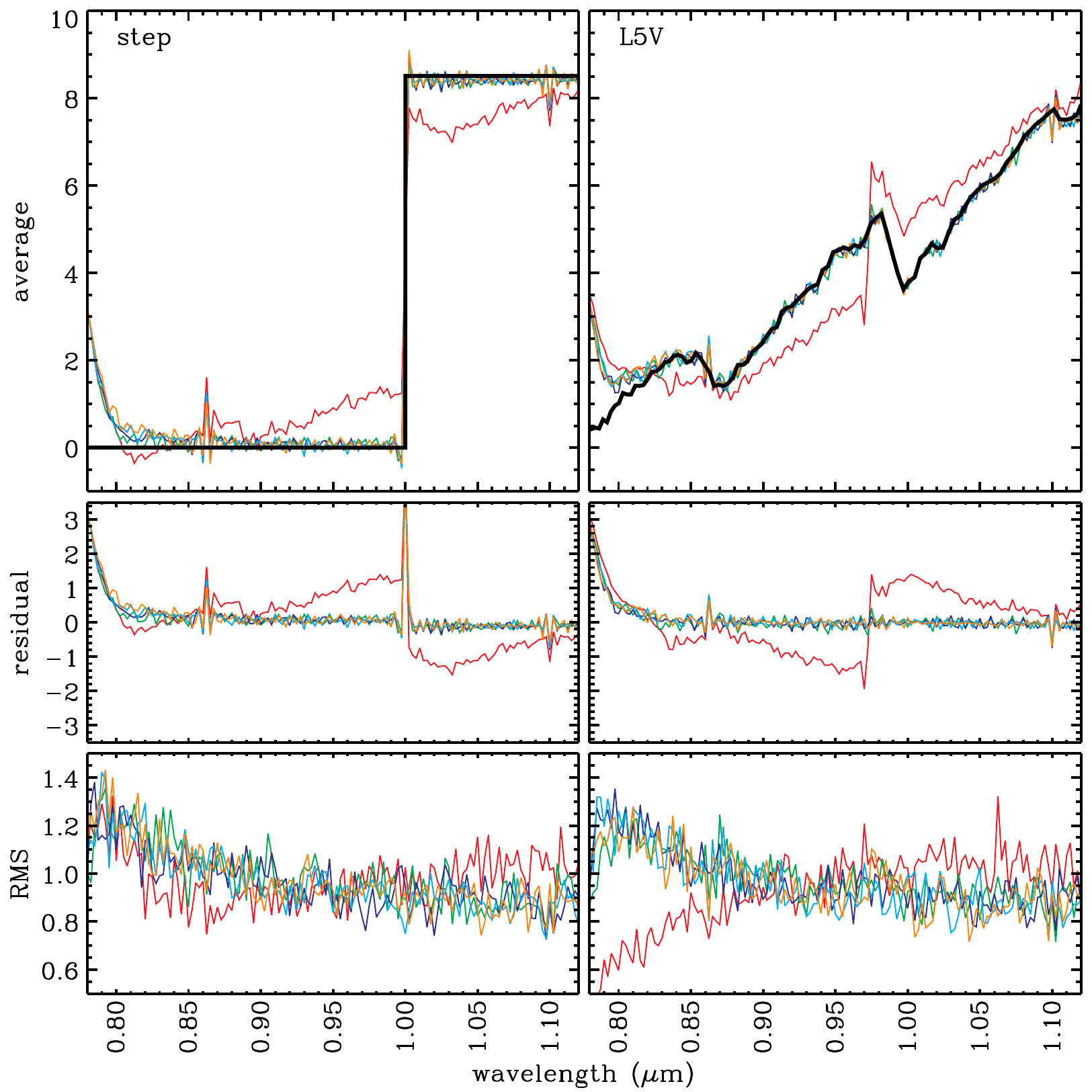}
  \caption{Spectral reconstruction of  two overlapping dispersions. As
    described in \Sect{sec:multiple}, we  consider five scenarios that
    vary  the  degree  of  degeneracy  between  the  two  sources,  as
    modulated   by   the   distribution  of   position   angles   (see
    \Tab{tab:orients}).  In the top, middle, and bottom panels we show
    the average, residual, and  RMS (respectively), which are computed
    from  the 100  iterations (see  \Sect{sec:multiple}).  In  the top
    panels,  we indicate  the  input spectra  used  for computing  the
    residuals,   otherwise    the   line    colors   are    given   in
    \Tab{tab:orients}.  For  the maximally-degenerate  scenario (\#1),
    \LINEAR\ is  clearly incapable  of reconstructing the  spectra for
    the   sources    with   any   confidence.    In    contrast,   the
    partially-degenerate  scenarios   (\#3,  4,  5)  do   not  perform
    demonstrably worse  than the fully-non-degenerate  scenario (\#2),
    which we consider to be the null hypothesis.}\label{fig:orients}
\end{figure}

One of  the key goals  of \LINEAR\ is to  better address the  issue of
contamination  when  the trace  of  multiple  sources overlap  on  the
detector.   To detail  the  success  of \LINEAR\  in  this regime,  we
consider  a  Monte Carlo  simulation  similar  to those  described  in
previous sections.   Here we simulate two  sources with $Y\!=\!18$~mag
that are separated $1\farcs5$ and have  a step-like and L5V spectra as
described  above.  In  \Fig{fig:confusion},  we  show an  illustrative
example  of the  two  sources  observed in  two  position angles  (for
simplicity   we   plot   the   sources  here   with   constant   SEDs:
$f_{\lambda}\!=\!f_0$).  Based on this intrinsic scene, we create five
observational scenarios, each of which consists of four distinct grism
images.    We   dither   these    four   images   according   to   the
\texttt{WFC3-IR-DITHER-BOX-MIN}      pattern     \citep[Table      C.3
  of][]{handbook}.   Each scenario  has  a  different distribution  of
position angles  (PAs; see \Tab{tab:orients}) designed  to bracket the
primary possibilities:
\begin{description}
  \item[fully  degenerate] the  sources overlap  in all  spectroscopic
    images; such as the case  in single orient, but possibly dithered,
    data  \citep[\eg][]{atek10,brammer12}.  It  is worth  noting that,
    this situation  may also arise  in the case  of data taken  in two
    orients that are offset by $180^\circ$;
  \item[partially degenerate] the sources  overlap in some images, but
    are uncontaminated in others; such as  the case with data taken at
    multiple  orients, but  possibly with  dithers at  a given  orient
    \citep[\eg][]{grapes,figs}; and
  \item[non-degenerate]  sources overlap  in no  spectroscopic images;
    such  as the  case  of distinct  sources, to  be  considered as  a
    \emph{null hypothesis}.
\end{description}
For each simulated grism image, we include background noise consistent
with  a $1200$~s  exposure with  G102  on WFC3/IR,  complete with  the
effects described in  \Sect{sec:tests}.  We simultaneously reconstruct
the two  spectra from  all four images  (within a  given observational
scenario), and iterate this  process 100 times.  In \Fig{fig:orients},
we show the extracted spectra for the five scenarios, each of which is
averaged  over  the 100  iterations.   The  colors  of the  lines  are
described in \Tab{tab:orients}.

\LINEAR\ is quite adept at reconstructing the overlapping dispersions,
provided at  least some of  orients are uncontaminated.  In  fact, the
difference between the partially-degenerate  scenarios (\#3, 4, 5) and
non-degenerate  scenario  (\#2)  are relatively  minor.   Furthermore,
\LINEAR\  cannot  reconstruct  the  spectra  of  the  sources  in  the
fully-degenerate  scenario (\#1),  which is  to be  expected.  In  the
fully-degenerate  scenario, \LINEAR\  tends to  produce very  similar
spectra for the  two sources, which have features  that are indicative
of  both objects.   For example,  the step-like  and strong  molecular
absorption   of   the   intrinsic   spectra   are   evident   in   the
fully-degenerate  case.   This  highlights the  importance  of  having
multiple  orients,  however  more  work  is  needed  to  estimate  the
appropriate number  or distribution  of orients  for an  arbitrary (or
specified) scene.

\begin{deluxetable}{ccl}
  \tablecaption{Position Angles\label{tab:orients}}
  \tablehead{\colhead{scenario} & \colhead{PAs} & \colhead{color\tablenotemark{$\dagger$}}\\
   \colhead{} &\colhead{(deg)} & \colhead{}}
  \startdata
  1 & 0,0,0,0    & red \\
  2 & 0,30,60,90 & green \\
  3 & 0,0,90,90  & blue \\
  4 & 0,0,0,90   & cyan \\
  5 & 0,45,45,90 & orange
  \enddata
  \tablenotetext{\dagger}{The color as shown in \Fig{fig:orients}.}
\end{deluxetable}

\section{A Case Study: The Hubble Ultra-Deep Field}\label{sec:hudf}

To demonstrate  the power of  \texttt{LINEAR}, we present  the WFC3/IR
grism     spectroscopy    of     the    Hubble     Ultra-Deep    Field
\citep[HUDF;][]{beckwith}.  There  are two programs that  observed the
HUDF in  the G141 grism;  we briefly  summarize the key  properties of
these data  in \Tab{tab:hudf}.  We  process these data  using standard
algorithms implemented  by \texttt{CALWF3}\footnote{\texttt{CALWF3} is
  a  product  of  the  Space Telescope  Science  Institute,  which  is
  operated  by AURA  for  NASA.}, but  apply additional  time-variable
background sky subtraction as described by \citet{figs}.  However, the
G141  data can  have  an additional  time-variable spectral  component
coming  from  earth-glow  \citep[e.g.][but  also  discussed  above  in
  \Sect{sec:backsubtract}]{brammer14}.   We use  the contemporaneously
obtained pre-imaging in the F140W band to refine the astrometry of the
grism  images, and  provide the  direct  image for  the pixel  weights
(${\cal I}(x_0,y_0)$ in \Eqn{eqn:weight}).  However, one of the visits
for program~12177 required higher  than normal background corrections,
therefore, to  avoid contaminating this  case study, we  exclude these
four exposures ($\sim\!5320$~s) from our analysis.  This results in 36
individual  images,   obtained  over   three  orients,  for   a  total
integration  time  of  $\sim\!43.31$~ks.  However  with  the  relative
positions  and  field  rotations,  not every  source  has  this  total
exposure  time.  In  principle,  this  results in  $\sim\!4\times10^7$
independent    measurements,   since    the   WFC3/IR    detector   is
$1014\times1014$~pix$^2$.    However  the   number   in  practice   is
$\sim\!1.92\times10^6$, since  only a  fraction of the  useable pixels
contain source flux.

\begin{deluxetable}{ccc}
  \tablecaption{HUDF Grism Observations\label{tab:hudf}}
  \tablehead{\colhead{PropID} & \colhead{\texttt{ORIENTAT}\tablenotemark{$\dagger$}} & \colhead{Exp.~Time}\\
    \colhead{} & \colhead{(deg)} & \colhead{(ks)}}
  \startdata
  12099 & $-178$ & $15.04$\\
  12177 & $+176$ & $18.85$\\
        & $+127$ & $9.42$
  \enddata
  \tablenotetext{\dagger}{As defined in the \texttt{fits} header.}
\end{deluxetable}

We    extract    one-dimensional     spectra    for    sources    with
F140W$\!\leq\!26$~mag  \citep[as  defined  by an  isophotal  magnitude
  described in][]{figs}, which resulted in $1,112$~individual sources.
For        each        source,        we        extract        between
$1.0\!\leq\!\lambda\!\leq\!1.7~\mu$m     with     a    sampling     of
$50$~\AA\   ($120$   elements   per    source),   which   results   in
$\sim\!1.1\times10^5$   unique   spectral  elements\footnote{Not   all
  wavelengths for every  source are sampled by the grism  data, so the
  true number  of parameters is  not simply the  sum of the  number of
  spectral elements over  all sources.}  to be  determined by \LINEAR.
The  $W$-matrix has  $\sim\!3.82\times10^8$  non-zero  elements for  a
sparsity    of   $\sim\!0.017$\%    and    a    Frobenius   norm    of
$||W||_F\!=\!8.2\times10^4$ e$^-$~s$^{-1}$ per $10^{-17}$ erg s$^{-1}$
cm$^{-2}$  \AA$^{-1}$.  For  comparison, we  also extract  the sources
through    the    same    apertures    using    standard    techniques
\citep[e.g.][]{figs}, and  then average combine  with inverse-variance
weights the three orients for each source.

\begin{figure}
  \includegraphics[width=0.47\textwidth]{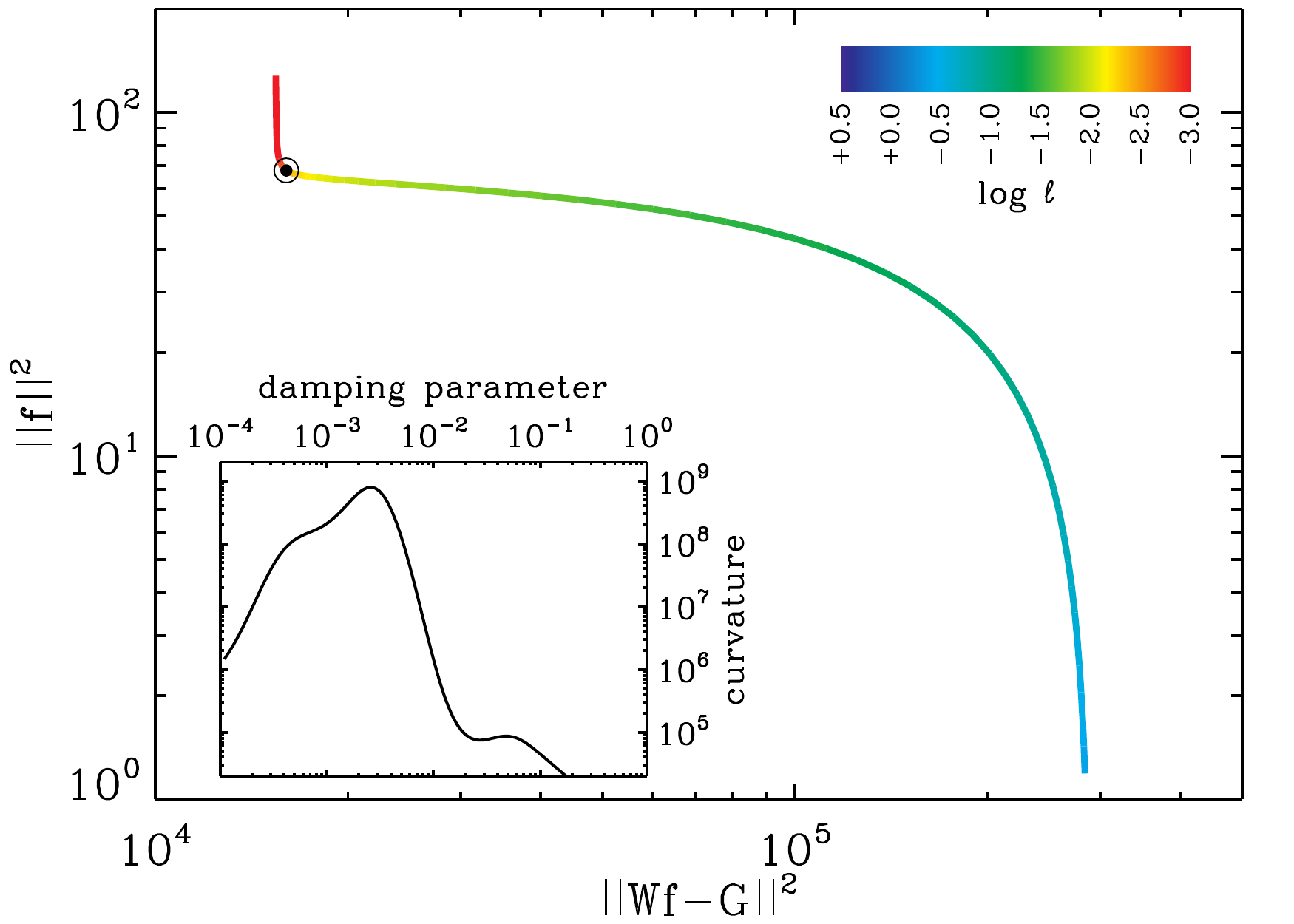}
  \caption{The  L-curve for  the G141  spectroscopy from  the HUDF  as
    described by  \citet{hansen}.  This analysis provides  a mechanism
    for  selecting  the  optimal  damping parameter  as  a  compromise
    between  modeling  the  data  and fluctuations  in  the  resulting
    spectra,  as   indicated  by   the  abscissa  and   the  ordinate,
    respectively.   The color  of  the line  indicates the  parametric
    value of the  damping, as indicated by the color  bar in the upper
    right  corner.  In  the  inset  to the  lower  left,  we show  the
    curvature as a function of the  damping that is used to select the
    optimal damping  parameter ($\ell_{\rm  opt}$), which  we indicate
    with the bullseye.  \label{fig:lcurve}}
\end{figure}

\begin{figure}
  \includegraphics[width=0.47\textwidth]{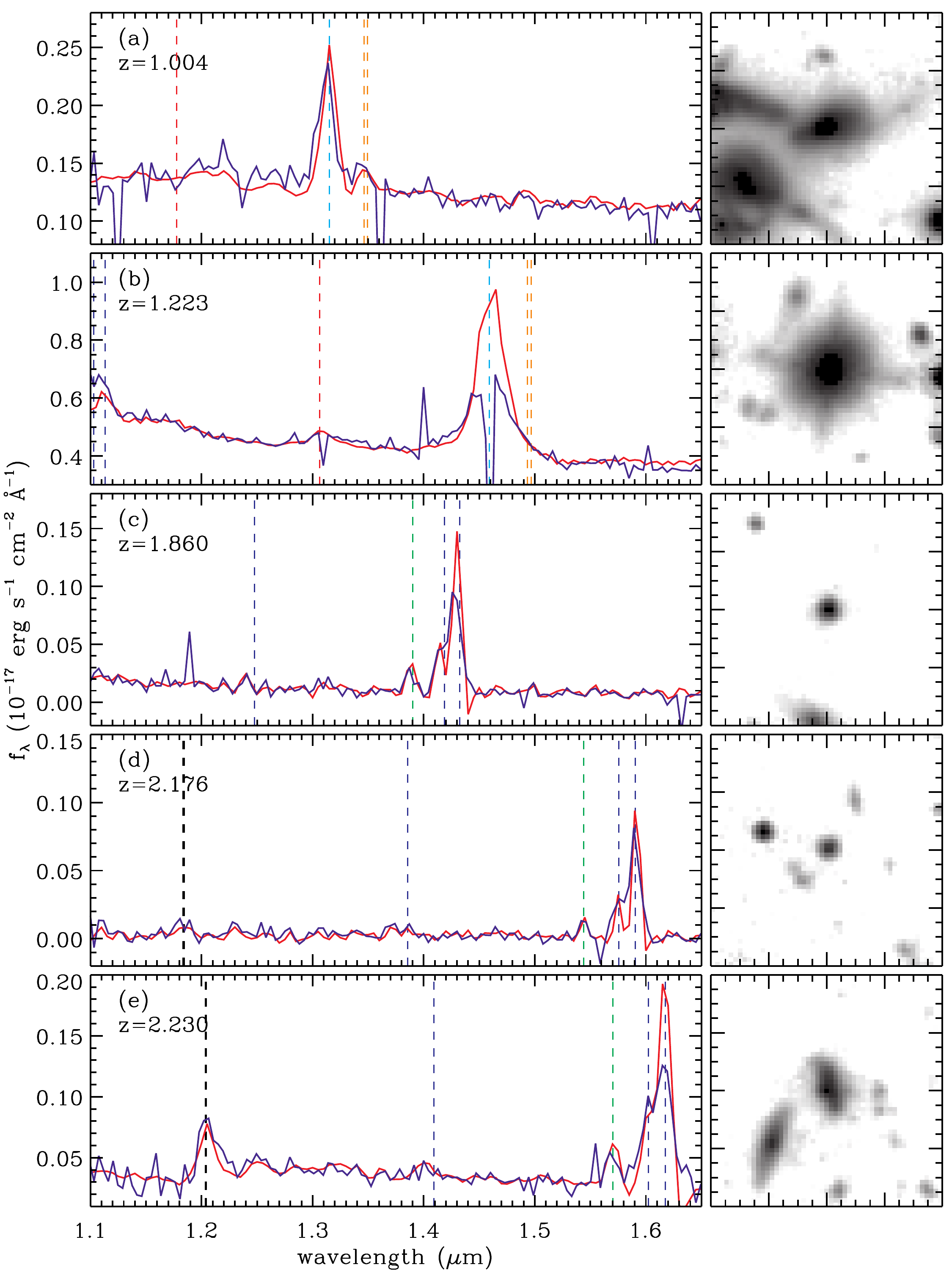}
  \caption{Illustrative   examples   of   \LINEAR\   from   the   G141
    spectroscopy  from the  HUDF arranged  by spectroscopic  redshift.
    The blue and red lines represent the standard, orient-averaged and
    \LINEAR\ spectra,  respectively.  Along the right  column, we show
    the F140W image used to establish  the pixel weights.  We show the
    redshifted wavelengths of common optical emission lines as colored
    vertical   dashed  lines,   but  the   most  common   species  are
    [\ion{O}{2}]  (black),  H$\beta$   (green),  [\ion{O}{3}]  (blue),
    \ion{He}{1}  (red), H$\alpha$  (cyan), and  [\ion{S}{2}] (orange).
    In \Fig{fig:zoom}, we  show zoomed versions of panels  (b) and (d)
    for clearer illustrations.\label{fig:comparison}}
\end{figure}

We  show the  L-curve  (as  \Sect{sec:damp}) for  these  G141 data  in
\Fig{fig:lcurve}, and the color of the line indicates the value of the
damping  parameter, as  shown in  the color  bar.  Although  there are
algorithms  for efficiently  locating the  point of  maximum curvature
\citep[e.g.][]{cult},  we  simply sweep  through  a  range of  damping
parameters to compute the point of  maximum curvature, as shown in the
inset to  \Fig{fig:lcurve}.  From  this we  find an  optimized damping
parameter  of  $\ell_{\rm  opt}\!=\!2.57\times10^{-3}$, which  is  the
value we adopt for  all subsequent analyses.  In \Fig{fig:comparison},
we show several  sources with strong emission lines,  whose colors are
described in the caption.

\begin{figure}
  \includegraphics[width=0.47\textwidth]{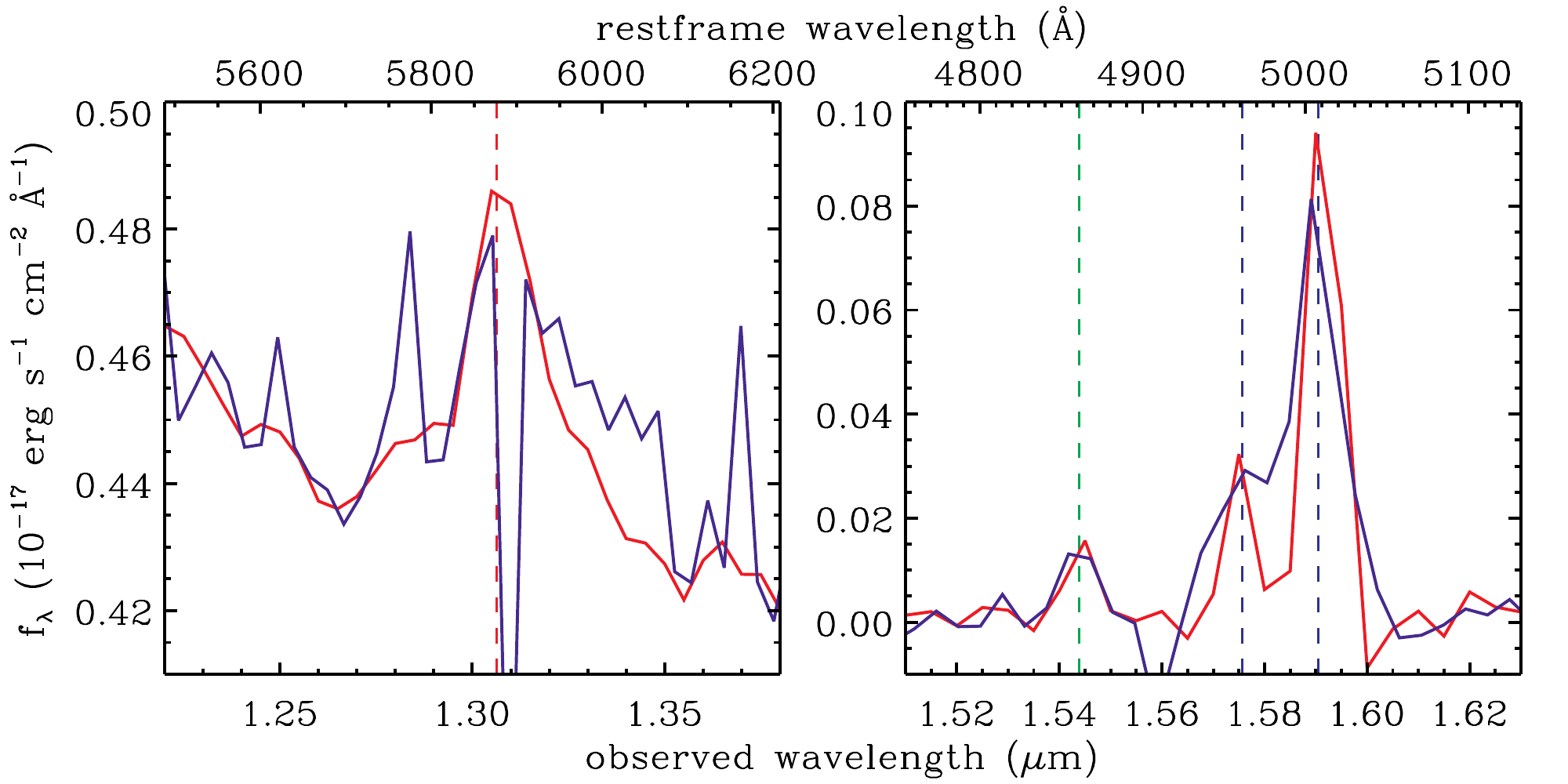}
  \caption{Zoomed    plots    of    panels   (b)    and    (d)    from
    \Fig{fig:comparison}.   The  colors  and  symbols  have  the  same
    meaning as in \Fig{fig:comparison}.   Here the shape and structure
    of the  \ion{He}{1} (left)  and the  H$\beta$/[\ion{O}{3}] complex
    (right) is seen in better detail, which reinforces the advantages
    discussed in \Sect{sec:hudf}. \label{fig:zoom}}
\end{figure}

Based on  these data in the  HUDF, we identify multiple  ways in which
\LINEAR\ improves  upon the  orient-averaged extractions.   First, the
spectral resolution for the \LINEAR\ spectra is $\sim\!20-30\%$ higher
than the  averaged spectra.   This improvement  is most  pronounced in
panels  (c),   (d),  and   (e)  of  \Fig{fig:comparison},   where  the
[\ion{O}{3}] and H$\beta$ complex is  very clearly resolved (see right
panel  of \Fig{fig:zoom}  for a  better representation).   This effect
arises for two reasons.  First,  the achievable spectral resolution is
limited by  the spatial profile  projected along the  dispersion axis.
But since  there are several  orients, which generally  have different
projected  spatial extents,  the individually  extracted spectra  will
have   different  effective   resolutions.   Therefore   the  weighted
averaging  increases the  overall  signal-to-noise at  the expense  of
degrading  the   high-resolution  spectra  from   the  high-resolution
orients.  Second, \LINEAR\ projects and deforms pixels from the direct
image  to the  set of  grism  images as  a function  of wavelength  to
compute the  fractional pixel areas.   Since each of the  grism images
have a  unique dither  position and orientation,  they provide  a more
complete spectral sampling.  This is  in analogy to drizzling dithered
images       to       improve       the       spatial       resolution
\citep[e.g.][]{fruchter,anton}.  As a final  note, the net improvement
in  the spectral  resolution depends  on  the geometry  of the  scene,
source(s), and observation(s), therefore  we expect these improvements
will vary from dataset to dataset.

A  second advantage  of \LINEAR\  concerns the  signal-to-noise, which
improves  by  $\sim\!20-30\%$.   For  example,  the  object  shown  in
\Fig{fig:comparison}(b)  or left  panel of  \Fig{fig:zoom} is  a known
quasar based  on the X-ray  flux and spectroscopic  redshift \citep[of
  $z\!=\!1.22$;][]{xu,xue}.  So  here the improvement to  the spectral
resolution  is less  profound, however  the increased  signal-to-noise
makes the  \ion{He}{1} emission at  $\lambda_{\rm obs}\!=\!1.304~\mu$m
easily detectable.  \LINEAR\ provides these  gains since we model each
grism image simultaneously, as opposed to stacking grism images of the
same orient (but dithered).  This improves the statistics by providing
more independent measurements of the underlying signal.

Finally,  the  first  source  in  \Fig{fig:comparison}(a)  shows  both
improvements   in  signal-to-noise   and   spectral  resolution,   but
importantly  this source  is  member  of a  very  complex grouping  of
interacting galaxies.  Consequently,  it has significant contamination
from  the neighbors,  however \LINEAR\  is able  to faithfully  remove
these extraneous signals.

\section{Subjects for Future Study} \label{sec:mods}

As  demonstrated,  \LINEAR\  is  already quite  powerful,  however  we
foresee several modifications in the near future.
\begin{enumerate}
  \item The current implementation of \LINEAR\ is highly geared toward
    WFC3/IR, however it would be straight-forward to include the other
    instruments on  HST and the  planned missions of JWST  and WFIRST.
    At  present,  \LINEAR\  is  only capable  of  working  with  grism
    configurations  in   the  \aXe-based  format.   As   described  in
    \Sect{sec:primer}, modifications  may be implemented to  work with
    the   generalized  transformations   (see  \Eqn{eqn:genrep})   for
    instruments   with  row/column   dispensers  \citep[as   with  the
      instruments  with \emph{The  James  Webb  Space Telescope},  see
    ][]{pr17} or  with multiple  detectors, particularly if  the trace
    from a  single order extends  on multiple detectors (such  as with
    WFIRST).
  
  \item It is important to  remove the astrophysical background light,
    and  as \citet{brammer_isr}  describe, this  can be  a non-trivial
    process.   However, the  flux coming  from the  background can  be
    considered  as  an additional  source  with  a distinct  spectrum.
    Therefore   it  is   formally  possible   to  perform   background
    subtraction \emph{within} the \LINEAR\ methodology, at the expense
    of making the $W$-matrix considerably denser in some rows/columns.
    A first step  is to implement a gray background  spectrum for each
    grism exposure, which would add little complexity, but may improve
    the  reconstruction  of  faint   sources  that  have  brightnesses
    comparable to the background errors.

  \item   We  have   framed  the   problem  of   reconstructing  grism
    spectroscopy as solving a system  of sparse, linear equations, and
    therefore the optimal set of source spectra can be easily obtained
    by standard computational techniques.  These obvious advantages in
    solving for  the ideal  non-parametric spectra  are offset  by the
    introduction of a  damping parameter.  However it  may be possible
    to  introduce a  limited  number of  non-linear parameters,  which
    would  likely modulate  global  properties  of the  reconstruction
    (such  as  something  that  governs  the  sky  background).   This
    additional  complexity   may  require  subsuming   the  LSQR-based
    minimization in a  single step of a non-linear  optimizer (such as
    Levenberg-Marquardt),   which   solves    for   these   additional
    parameters.

  \item It may  be additionally possible to restrict the  scope of the
    LSQR-based  matrix  solution  by  only  considering  sources  that
    overlap  in the  dispersed  images.  This  would  be analogous  to
    solving  a  \emph{friends-of-friends}-type   problem,  where  each
    separate set of friends is  solved in parallel.  This approach may
    be implemented by matrix operations on $W$.

  \item  Because  WFC3/IR  non-destructively samples  the  reads,  any
    incident cosmic  rays can  be flagged  by linear  regression.  The
    current WFC3/IR calibration software implicitly assumes that there
    are no time-dependent signals (such  as the sky background varying
    with  time), which  can easily  confound the  cosmic ray  flagging
    \citep[\eg][]{brammer_isr,bram16,pirzkal17}.         Additionally,
    optical detectors (such  as the Advanced Camera  for Surveys) will
    not have  the non-destructive sampling and,  therefore may require
    additional processing  steps in  the reconstruction to  remove the
    cosmic rays.

\end{enumerate}

\section{Discussion}\label{sec:discuss}

The \LINEAR\  algorithm leverages all  of the spectral  information at
the reconstruction phase to  produce the best one-dimensional spectrum
consistent with  the data.   We foresee  several situations  where the
\LINEAR\ paradigm will be a major improvement over standard algorithms
for analyzing slitless spectroscopy.

\begin{description}
  \item[crowded fields] Standard extraction  tools simply flag regions
    of a two-dimensional dispersed  image that have contributions from
    multiple sources.   In a  future work, we  present a  use-case for
    spectroscopy of high-redshift sources  lensed by foreground galaxy
    clusters  (Ryan et  al.~2018, in  preparation).  \LINEAR\  may be
    equally  adept at  spectroscopy in  dense stellar  fields, however
    further tests may be necessary  to establish the number of orients
    needed for complete spectroscopic coverage.

 \item[spatially resolved spectroscopy] We  have framed the notion of
   a  \emph{source}  as a  collection  of  pixels  assumed to  have  a
   consistent spectral  shape.  Importantly, these collections  may be
   associated with a single physical object, such as a large, resolved
   galaxy or nebulosity.  Therefore it  may be possible to partition a
   single astrophysical object into many sources, and \LINEAR\ will be
   capable of extracting individual one-dimensional spectra separately
   for each partition.  In this way, \LINEAR\ may approximate slitless
   spectroscopy as  an integrated-field unit (IFU).   However, it will
   be important  to keep certain  statistical issues in mind,  such as
   the  relative  numbers of  knowns  and  unknowns (as  discussed  in
   \Sect{sec:code}).

 \item[strict sampling requirements] It may  be that a survey has very
   strict requirements  on any completeness and/or  reliability of the
   extracted spectroscopy or the  derived products (such as redshift).
   For example,  the various  cosmology experiments with  WFIRST place
   stringent expectations  on the spectroscopic redshift  accuracy and
   completeness \citep[or sample size;][]{wfirst}.  Furthermore strong
   positional biases in  the redshift quality, whether  they stem from
   the  quality of  existing photometry  to make  a crude  estimate of
   photometric       redshift        \citep[\eg][]{brammer12}       or
   confusion/contamination  from  crowded  regions,  is  important  to
   characterize  and resolve.   Therefore  an entirely  self-contained
   framework  that   does  not   rely  on  existing   photometry  like
   \LINEAR\ will be an important analysis technique.
   
\end{description}

\section{Summary}\label{sec:summary}

We   have  presented   a   new  algorithm   for  extracting   slitless
spectroscopy, and  demonstrated the success with  archival HST WFC3/IR
data.   This   algorithm  was   devised  to  explicitly   correct  for
overlapping  dispersions and  circumvent  the  need for  contamination
corrections. In forthcoming works, we will show additional  use cases,
particularly regarding the mitigation of contaminating sources.

\acknowledgments

We would like to thank  W.~Landsman, C.~Markwardt, and J.~D.~Smith for
their kind permission to use  and redistribute their programs.  We are
very grateful  to J.~MacKenty and  E.~Sabbi for advice and  support in
this  endavour.   We also  appreciate  many  vibrant discussions  with
N.~Grogin, V.~Dixon,  G.~Brammer, M.~Sosey,  J.~Colbert, A.~Koekemoer,
S.~Malhotra, J.~Rhoads,  K.~Gilbert, L.~Smith,  R.~van der  Marel, and
L.~Armus.  We  thank G.~Schwarz for  his assistance in  preparing some
components  of this  \LaTeX\ document.   This research  has benefitted
from  the  SpeX  Prism  Library,   maintained  by  Adam  Burgasser  at
\href{http://www.browndwarfs.org/spexprism}{http://www.browndwarfs.org/spexprism}.
We  thank the  anonymous  Referee for  the  excellent suggestions  and
constructive  feedback.   Part  of  this work  carried  out  at  Space
Telescope Science  Institute (STScI) was  performed in the  context of
Science Operations Center  contracts for the HST,  and WFIRST missions
funded by  NASA Goddard  Space Flight  Center.  Based  on observations
made with the NASA/ESA Hubble  Space Telescope, obtained from the data
archive at the Space Telescope Science Institute. STScI is operated by
the Association of Universities for  Research in Astronomy, Inc. under
NASA contract NAS 5-26555.  

\vspace{5mm}
\facilities{HST(WFC3)}

\software{IDL, C}

\appendix

\section{Notation Glossary}
To avoid any confusion with  the above notation, we provide a glossary
to define the key variables and their units.

\begin{deluxetable}{llp{3in}}
\tablecaption{Notation Glossary\label{tab:glossary}}
\tablehead{\colhead{symbol} & \colhead{units} & \colhead{explanation}}
\startdata
${\cal G}_i$ & & the \ith\ grism image\\
${\cal G}_{x,y,i}$ or ${\cal G}_{\vartheta}$ & e$^-$~s$^{-1}$ & the measured brightness at pixel $(x,y)$ in the \ith\ grism image\\
${\cal U}_{x,y,i}$ or ${\cal U}_{\vartheta}$ & e$^-$~s$^{-1}$ & the measured uncertainty at pixel $(x,y)$ in the \ith\ grism image\\
$\tilde{y}(\tilde{x})$  & pixel & spectral trace in relative coordinates\\
$S(\lambda)$ & e$^-$~s$^{-1}$  per erg~s$^{-1}$~cm$^{-2}$~\AA$^{-1}$& field-averaged sensitivity as a function of wavelength\\
${\cal F}(x,y;\lambda)$ & dimensionless &  flat-field cube\\
${\cal  F}_j(x,y)$ & dimensionless & the \jth\ component for the flat-field cube coefficients\\
${\cal I}_i(x,y)$ & dimensionless & normalized, direct image of the \ith\ source\\
$f_{\lambda,j}$ or $f_{\varphi}$ & $10^{-17}$~erg~s$^{-1}$~cm$^{-2}$~\AA$^{-1}$ & calibrated spectrum for the \jth\ source\\
$f$ & $10^{-17}$~erg~s$^{-1}$~cm$^{-2}$~\AA$^{-1}$ & collection of computed spectra --- the same as $f_{\lambda,i}$ but \emph{flattened} into a single array\\
$\hat{f}$ &  $10^{-17}$~erg~s$^{-1}$~cm$^{-2}$~\AA$^{-1}$ & optimized solution for the set of source spectra\\
$f_0$ & $10^{-17}$~erg~s$^{-1}$~cm$^{-2}$~\AA$^{-1}$ & collection of damping targets --- the same dimensionality as $f$\\
$s(\tilde{x})$ & pixel & path length along the spectral trace\\
$(x,y)$  & pixel & position on the detector\\
$(x_0,y_0)$ & pixel & undispersed position of a source/pixel\\
$(x_\mathrm{off},y_\mathrm{off})$ & pixel & shift in detector coordinates between undispersed position and start of the spectral trace\\
$\delta\lambda$ & \AA\ & subsampling bandwidth interval for propagation of area compuations\\
$\Delta\lambda_i$ & \AA\ & extraction bandwidth for the \ith\ source\\
$\mathrm{d}\lambda/\mathrm{d}s$ & \AA~pixel$^{-1}$ & native spectral resolution of the detector\\
$\ell$ & dimensionless & damping parameter for the damped, least-squares solution\\
$W_{\vartheta,\varphi}$ or $w$ & $10^{-17}$~erg~s$^{-1}$~cm$^{-2}$~\AA$^{-1}$ per e$^-$~s$^{-1}$ & matrix element to convert between unknowns (spectra) and knowns (pixel brightnesses)\\
$a(\lambda)$ & dimensionless & fractional pixel area between the direct image and a grism image at wavelength $\lambda$
\enddata
\end{deluxetable}

%
%

\end{document}